\documentclass[preprint,12pt]{elsarticle}

\usepackage{graphicx}
\usepackage{amssymb}
\usepackage{amsthm}

\journal{Journal of Computational Physics}

\begin{document}

\begin{frontmatter}

\title{Numerical modeling of transient two-dimensional viscoelastic waves}

\author[LMA]{Bruno Lombard\corref{cor1}}
\ead{lombard@lma.cnrs-mrs.fr}
\author[LMA]{Jo\"{e}l Piraux}
\ead{piraux@lma.cnrs-mrs.fr}
\cortext[cor1]{Corresponding author. Tel.: +33 491 16 44 13.}
\address[LMA]{Laboratoire de M\'{e}canique et d'Acoustique, 13402, Marseille, France}

\begin{abstract}
This paper deals with the numerical modeling of transient mechanical waves in linear viscoelastic solids. Dissipation mechanisms are described using the generalized Zener model. No time convolutions are required thanks to the introduction of memory variables that satisfy local-in-time differential equations. By appropriately choosing the relaxation parameters, it is possible to accurately describe a large range of materials, such as solids with constant quality factors. The evolution equations satisfied by the velocity, the stress, and the memory variables are written in the form of a first-order system of PDEs with a source term. This system is solved by splitting it into two parts: the propagative part is discretized explicitly, using a fourth-order ADER scheme on a Cartesian grid, and the diffusive part is then solved exactly. Jump conditions along the interfaces are discretized by applying an immersed interface method. Numerical experiments of wave propagation in viscoelastic and fluid media show the efficiency of this numerical modeling for dealing with challenging problems, such as multiple scattering configurations.
\end{abstract}


\end{frontmatter}



\section{Introduction}\label{SecIntro}

Wave motion in real media differs in many aspects from motion in an idealized elastic medium. The dispersion and attenuation induced, for instance, by grain-to-grain friction can greatly affect the amplitude of the waves and their arrival times. In the case of small perturbations, linear models of viscoelasticity provide reasonably accurate means of describing these effects. Viscoelastic constitutive laws give the stress in terms of the past strain rate history. 

The aim of this paper is to simulate the propagation and the diffraction of transient viscoelastic waves. We propose a new approach in three steps:
\begin{description}
\item[(i)] The generalized Zener model is used \cite{CARCIONE07}. The convolution products are then replaced by a set of local-in-time differential equations coupled with the evolution equations of velocity and stress. Moreover, usual attenuation laws can be approximated closely.
\item[(ii)] The evolution equations are splitted into two parts: a propagative part, which is solved using a fourth-order finite-difference ADER scheme on a Cartesian grid \cite{SCHWARTZKOPFF04}; and a diffusive part, which is solved analytically. Doing so ensures an optimal condition of stability. 
\item[(iii)] The jump conditions along the interfaces are discretized by an immersed interface method, which introduces a subcell resolution of the geometry and maintains the convergence rate of the scheme despite the non-smoothness of the solution. See \cite{LI06} for an overview of these methods. 
\end{description}
The generalized Zener model (i) has been addressed by various means, such as finite difference methods \cite{ROBERTSSON94,XU95}, spectral methods \cite{CARCIONE93}, spectral-element methods \cite{KOMATITSCH99}, finite element methods \cite{GROBY06,BECACHE04}, to cite only a few. The steps (ii) and (iii) combine the computational efficiency of Cartesian grid methods and an accurate description of the interfaces, as stated in the case of non-dissipative media \cite{LOMBARD04,LOMBARD06} and applied to computationally challenging configurations \cite{CHEKROUN09}.

The article is organized as follows. In section \ref{SecPb}, the generalized Zener model is presented; the method used to determine its parameters to simulate a given quality factor is also described. In section \ref{SecIbvp}, the evolution equations are written in the form of a first-order hyperbolic system with a source term; the jump conditions along the interfaces are also stated. The numerical methods used are introduced in section \ref{SecMethods}, including the numerical scheme and the splitting for the integration of the evolution equations, and the immersed interface method for the discretization of the jump conditions. Numerical experiments are presented in section \ref{SecNum} in the case of a viscoelastic / fluid interface. Comparisons with analytical solutions are proposed. A numerical experiment involving multiple scattering in a random medium also confirms the efficiency of the approach. Lastly, the perspectives are discussed in section \ref{SecPerspective}.


\section{Physical modeling}\label{SecPb}

\subsection{Constitutive law}\label{SecPbZener}

In a viscoelastic solid undergoing small perturbations, the stress depends linearly on the history of the past strain rates. In 1D, one writes
\begin{equation}
\sigma=\psi*\frac{\textstyle \partial\,\varepsilon}{\textstyle \partial\,t},
\label{Boltzmann}
\end{equation}
where $\sigma$ is the stress, $\varepsilon=\frac{\partial\,u}{\partial\,x}$ is the strain, $u$ is the displacement, $\psi(t)$ is the relaxation function, and $*$ denotes the time convolution. 

Various models of viscoelasticity can be found in the literature \cite{CARCIONE07}. The {\it Maxwell model} predicts a vanishing asymptotic residual stress. It therefore appears more appropriate for representing viscoelastic fluids. The {\it Kelvin-Voigt model} is computationally advantageous \cite{CARCIONE04}, but it predicts an unbounded phase velocity as frequency increases. Here, we choose the {\it generalized Zener model}, which accurately mimics the mechanical behavior of classical viscoelastic media during relaxation experiments:
\begin{equation}
\psi(t)=\pi_r\left(1+\sum_{\ell=1}^{N_r} \kappa_\ell\, e^{-\theta_\ell\,t}\right)\,H(t),
\label{Psi}
\end{equation}
where $H$ refers to the Heaviside distribution, $N_r$ is the number of relaxation mechanisms, $\theta_\ell$ are relaxation frequencies, and the coefficients $\kappa_\ell$ are strictly positive. The instantaneous unrelaxed state is denoted by $\pi_u$, and at the end of the process, the relaxation function has returned completely to the positive relaxed modulus $\pi_r$, where $0<\pi_r<\pi_u$. The phase velocity increases with the frequency, from $c_0=\sqrt{\pi_r/\rho}$ at null frequency to $c_\infty=\sqrt{\pi_u/\rho}$ at infinite frequency, where $\rho$ is the density \cite{CARCIONE07}. 


\subsection{Determination of the parameters}\label{SecPbEstime}

Let ${\cal F}$ be the Fourier transform of a function $g(t)$
\begin{equation}
{\cal F}\left(g\right)=\int_{-\infty}^{+\infty}g(t)\,e^{-\,i\,\omega\,t}\,dt,
\label{TFourier}
\end{equation}
where $\omega$ is the angular frequency. From (\ref{Psi}), the modulus of viscoelasticity $M(\omega)={\cal F}(\frac{\partial\,\psi}{\partial\,t})$ is:
\begin{equation}
M(\omega)=\pi_r\left(1+i\,\omega\sum_{\ell=1}^{N_r}\frac{\textstyle \kappa_\ell}{\textstyle \theta_\ell+i\,\omega}\right),
\end{equation}
and the ratio between the imaginary and real parts of $M$ is:
\begin{equation} 
Q^{-1}(\omega)=\frac{\textstyle \displaystyle \sum_{\ell=1}^{N_r} \frac{\textstyle \omega\,\theta_\ell\,\kappa_\ell}{\textstyle \theta_\ell^2+\omega^2}}{\textstyle 1+ \displaystyle\sum_{\ell=1}^{N_r} \frac{\textstyle \omega^2\,\kappa_\ell}{\textstyle \theta_\ell^2+\omega^2}}.
\label{Qm1}
\end{equation}
The quality factor $Q$ characterizes the attenuation of the viscoelastic waves.

To determine the $2\,N_r$ coefficients $\kappa_\ell$ and $\theta_\ell$ in (\ref{Psi}), we choose to minimize the distance between $Q^{-1}(\omega)$ and a given $Q_{ref}^{-1}(\omega)$ in a band of angular frequencies $[\omega_0,\,\omega_1]$. Here we have implemented a classical linear least-squares minimization procedure in the $L_2$ norm \cite{DAY84,EMMERICH87,BLANCH95}. Relaxation frequencies are distributed linearly on a logarithmic scale of $N_r$ points, ranging from $f_0=\omega_0\,/\,(2\,\pi)$ to $f_1=\omega_1\,/\,(2\,\pi)$ \cite{GROBY06}
\begin{equation}
\theta_\ell=\frac{\textstyle \omega_0}{\textstyle 2\,\pi}\,\left(\frac{\textstyle \omega_1}{\textstyle \omega_0}\right)^{\frac{\ell-1}{N_r-1}},\qquad \ell=1,...,\,N_r.
\label{Theta}
\end{equation}
The angular frequencies $\omega_0$ and $\omega_1$ obviously depend on the spectra of the source. The coefficients $\kappa_\ell$ are then obtained by solving the over-determined linear system deduced from (\ref{Qm1})
\begin{equation}
\sum_{\ell=1}^{N_r}\frac{\textstyle {\tilde \omega}_k\,\left(\theta_\ell-{\tilde \omega}_k\,Q_{ref}^{-1}({\tilde \omega}_k)\right)}{\textstyle \theta_\ell^2+{\tilde \omega}_k^2}\,\kappa_\ell=Q_{ref}^{-1}({\tilde \omega}_k), \qquad k=1,...,\,2\,N_r-1,
\label{Emmerich}
\end{equation}
where ${\tilde \omega}_k$ are distributed linearly on a logarithmic scale of $2\,N_r-1$ points
\begin{equation}
{\tilde \omega}_k=\omega_0\left(\frac{\textstyle \omega_1}{\textstyle \omega_0}\right)^{\frac{k-1}{2\,(N_r-1)}},\qquad k=1,...,\,2\,N_r-1.
\end{equation}
In our numerical experiments, calculations based on (\ref{Emmerich}) have never yielded non-physical negative values of $\kappa_\ell$, even with highly attenuating media (typically $Q_{ref}=5$). If necessary, more sophisticated methods can be applied. For instance, a nonlinear least-squares constraint optimization was used in \cite{BECACHE04} to ensure that the coefficients $\kappa_\ell$ were positive. An alternative method of optimization in the norm $L_\infty$ was presented in \cite{ASVADUROV04}, but this latter method is restricted to materials with constant $Q_{ref}$. 

To determine the number of relaxation mechanisms $N_r$, one can compare $Q_{ref}$ and the quality factor $Q$ deduced from (\ref{Qm1}) after optimization. However, the modeling error in the time domain is not easily deduced. A second idea consists in comparing the transient 1-D analytical solutions associated with $Q_{ref}$ and $Q$. These solutions are calculated using classical Fourier techniques, which are not described  here. 


\subsection{Numerical examples}\label{SecPbNum}

\begin{figure}[htbp]
\begin{center}
\begin{tabular}{cc}
(a) & (b)\\
\includegraphics[scale=0.31]{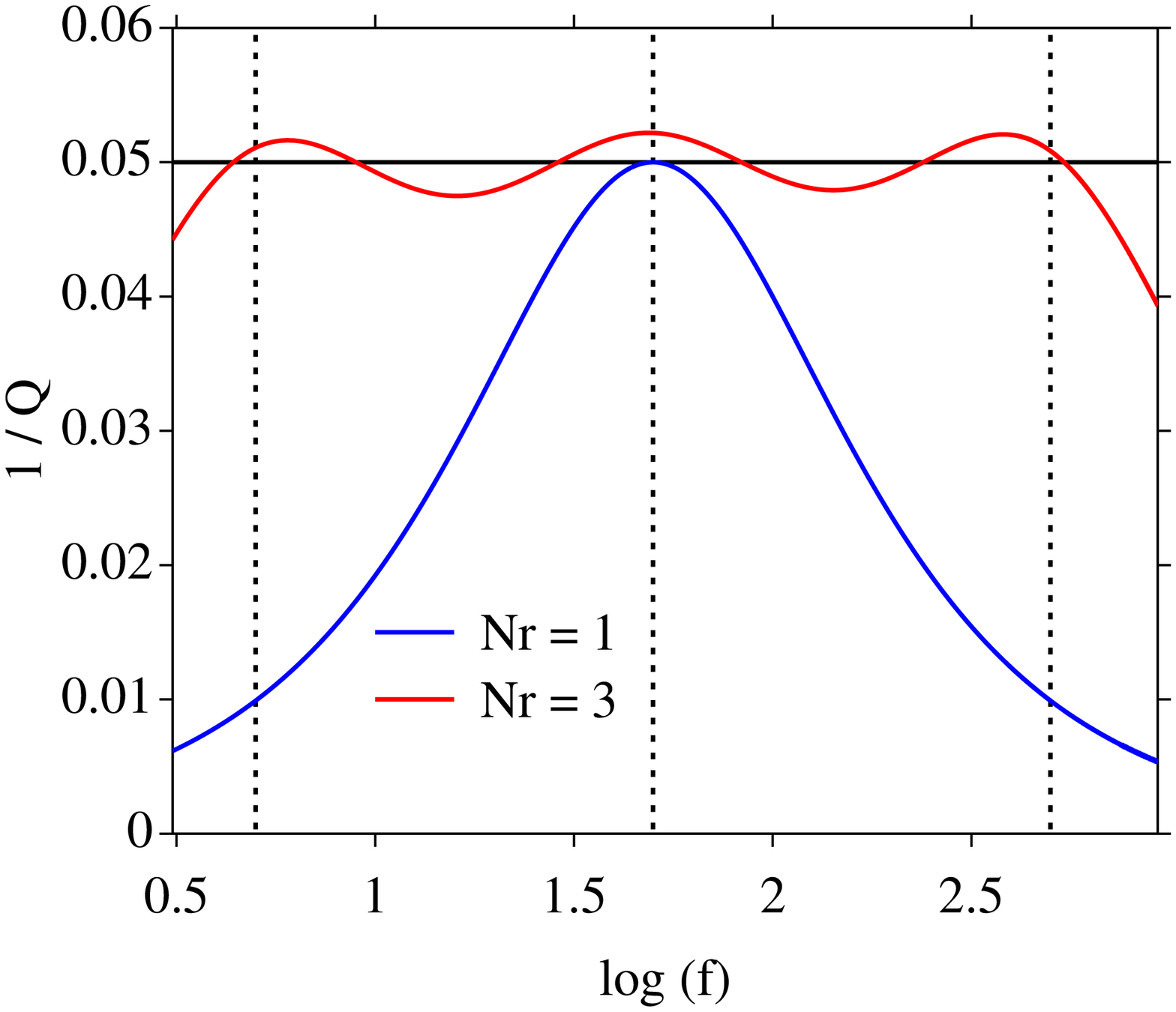}&
\includegraphics[scale=0.31]{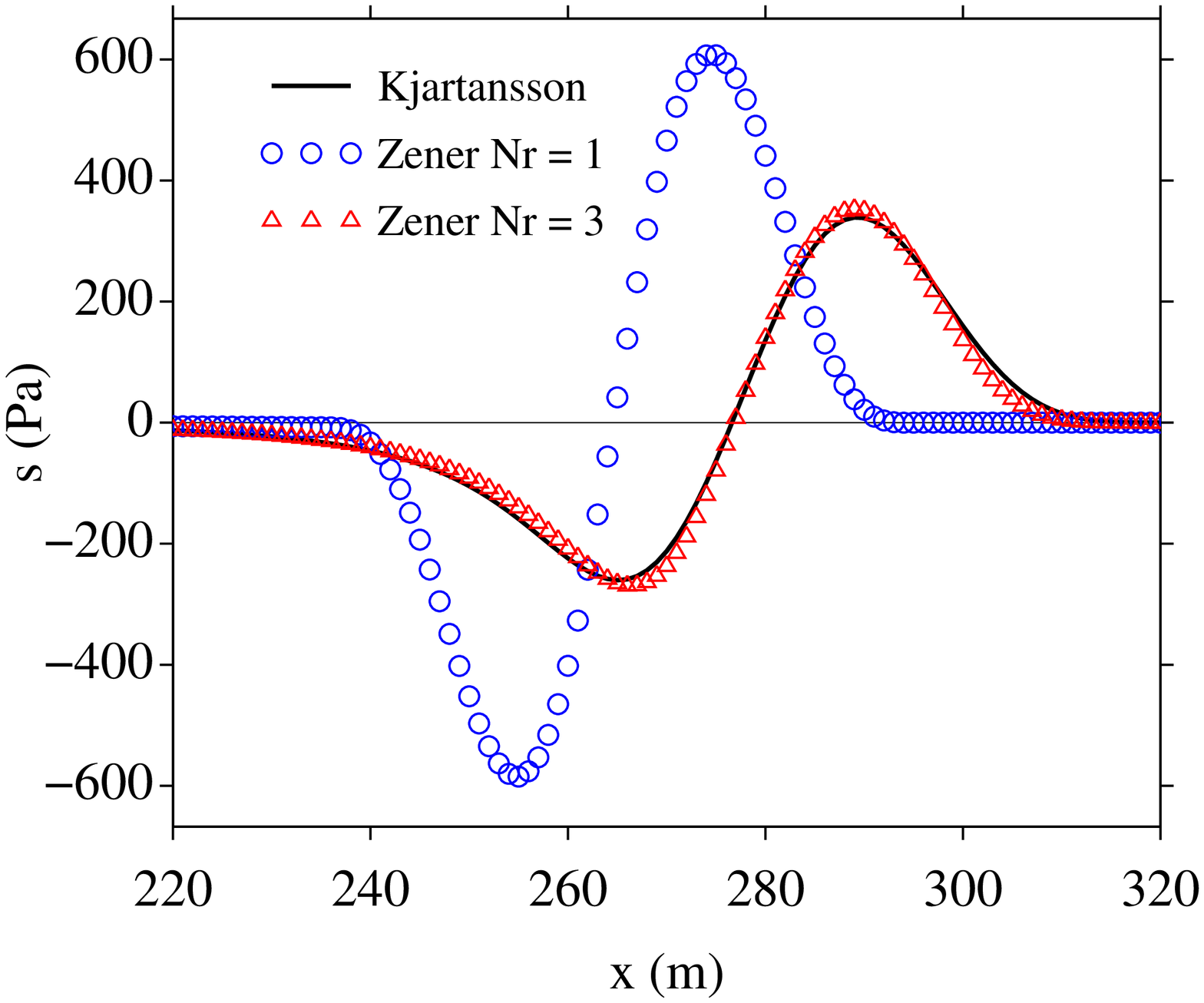}\\
(c) & (d)\\
\includegraphics[scale=0.31]{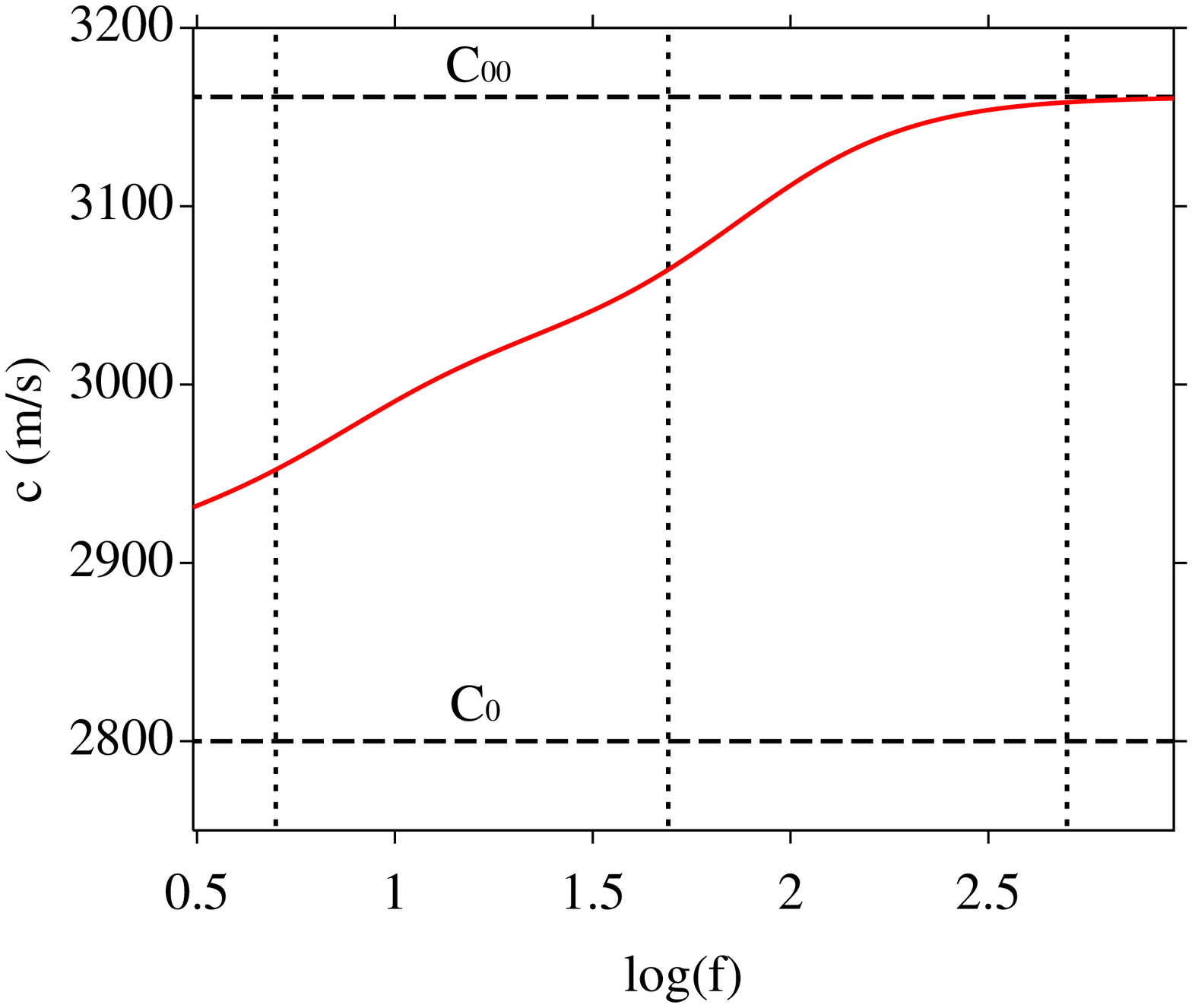}&
\includegraphics[scale=0.31]{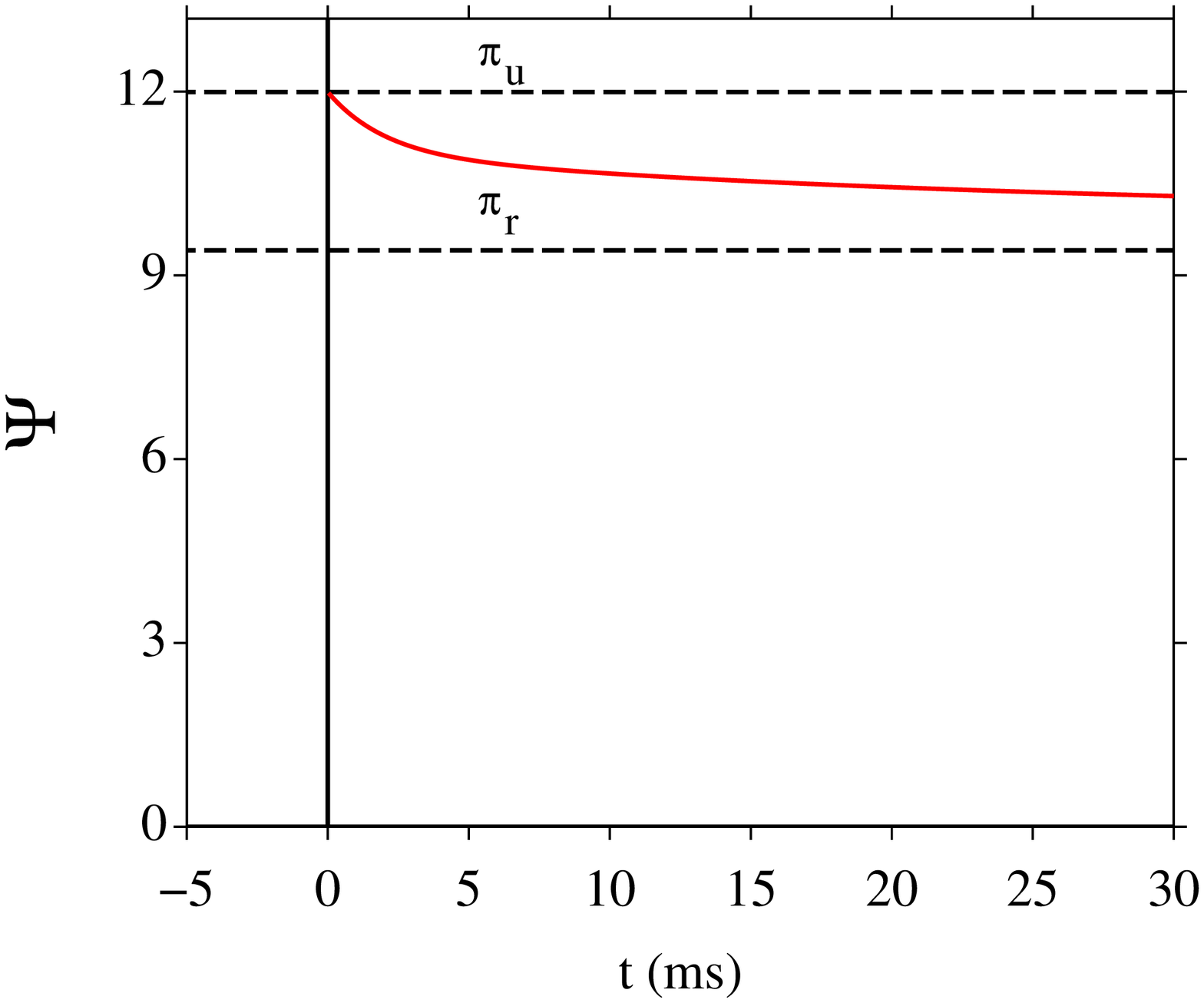}
\end{tabular}
\end{center}
\caption{Determination of viscoelastic parameters, with $Q_{ref}=20$, $f_c=50$ Hz, and $N_r=1$ or $N_r=3$ relaxation mechanisms. Quality factors (a); the solid horizontal line gives the exact value $1\,/\,Q_{ref}$. Time-domain 1-D analytical solutions obtained with the constant-Q model (Kjartansson's model) and the generalized Zener model (b). Phase velocity (c) and relaxation function (d) obtained with $N_r=3$. Physical parameters are those used in section \ref{SecNum}. In (a) and (c), the dotted vertical lines give the relaxation frequencies when $N_r=3$.}
\label{FigZener}
\end{figure}

The determination of the parameters in (\ref{Psi}) is illustrated in figure \ref{FigZener}. The set up is the same here as in section \ref{SecNum}: the source is a smoothly truncated sinusoid of central frequency $f_c=50$ Hz; the optimization of the quality factor is done between $f_0=f_c\,/\,10$ and $f_1=10\,f_c$; the physical parameters are $\rho=1200\,\mbox{ kg/m}^3$, $c_0=2800\,\mbox{ m/s}$, $Q_{ref}=20$. Considering a constant $Q_{ref}$ is usual in geosciences, where the real media show a quasi-constant quality factor within very wide frequency ranges \cite{AKI80}. In addition, the exact solution associated with a constant $Q_{ref}$ is particularly simple to obtain, involving the Kjartansson formula \cite{THESE_KJARTANSSON}. Note that non-constant $Q_{ref}$ can also be considered in numerical modeling without any restrictions.

Figure \ref{FigZener} compares the quality factor (a) and the time-domain exact solutions (b), when $N_r=1$ or $N_r=3$. Increasing $N_r$ clearly decreases the error introduced by describing the constant quality factor medium in terms of a finite number of relaxation mechanisms. At the same time, increasing $N_r$ greatly increases the computational cost, especially the memory requirements. In practice, $N_r>3$ is rarely implemented in the literature, and $N_r=1$ is widely used, especially in the 3-D context \cite{ROBERTSSON94}. 

Dispersion and relaxation curves shown in figure \ref{FigZener} (c) and (d) are computed taking $N_r=3$ relaxation mechanisms. One observes the predicted behavior: strict increase of the phase velocity from $c_0$ to $c_\infty$ (c), strict decrease of the relaxation function from $\pi_u$ to $\pi_r$ (d).


\section{Initial boundary-value problem}\label{SecIbvp}

\subsection{Constitutive law in two-dimensions}\label{SecIbvpZener}

The viscoelastic law (\ref{Boltzmann}) is generalized so that it applies to all space dimensions. In the 2-D case, the constitutive law governing a linear isotropic viscoelastic medium is \cite{CARCIONE07}
\begin{equation}
\sigma_{ij}=\left(\psi_{\pi}(t)-2\,\psi_{\mu}(t)\right)*\frac{\textstyle \partial\,\varepsilon_{kk}}{\partial\,t}\,\delta_{ij}+2\,\psi_\mu(t)*\frac{\textstyle \partial\,\varepsilon_{ij}}{\partial\,t},
\label{SigmaIJ}
\end{equation}
where $\sigma_{ij}$ and $\varepsilon_{ij}$ are the components of the stress and strain tensors, and $\delta_{ij}$ is the Kronecker symbol. With the generalized Zener model, the relaxation functions $\psi_\pi$ and $\psi_\mu$ are given by 
\begin{equation}
\begin{array}{l}
\displaystyle
\psi_\pi(t)=\pi_r\left(1+\sum_{\ell=1}^{N_r} \kappa^p_\ell\, e^{-\theta_\ell\,t}\right)\,H(t),\\
[10pt]
\displaystyle
\psi_\mu(t)=\mu_r\left(1+\sum_{\ell=1}^{N_r} \kappa^s_\ell\, e^{-\theta_\ell\,t}\right)\,H(t),\\
\end{array}
\label{Psi2D}
\end{equation}
where $\pi_r=\rho\,c_{p_0}^2$ and $\mu_r=\rho\,c_{s_0}^2$ are relaxed moduli under compressional and shear loads. The phase velocities of the compressional (P) and shear (S) waves at zero frequency are denoted $c_{p0}$ and $c_{s0}$. The unrelaxed moduli are written 
\begin{equation}
\displaystyle
\pi_u=\pi_r\left(1+\sum_{\ell=1}^{N_r}\kappa_\ell^p\right)=\rho\,c_{p_\infty}^2,\qquad
\mu_u=\mu_r\left(1+\sum_{\ell=1}^{N_r}\kappa_\ell^s\right)=\rho\,c_{s_\infty}^2,
\label{Cpinf}
\end{equation}
where $c_{p_\infty}$ and $c_{s_\infty}$ are the phase velocities of P and S waves at infinite frequency. The parameters $\theta_\ell$, $\kappa_\ell^p$ and $\kappa_\ell^s$ in (\ref{Psi2D}) are determined as in section \ref{SecPbZener} from the quality factors $Q^p_{ref}$ and $Q^s_{ref}$ of P and S waves. Usually, $Q^s_{ref}<Q^p_{ref}$: the S waves are more attenuated than the P waves. The relaxation frequencies $\theta_\ell$ are the same with both P and S waves, since they depend only on the frequency band of interest (\ref{Theta}). In addition, describing P and S waves with identical relaxation times, as well as identical numbers of relaxation mechanisms, greatly reduces the memory requirements \cite{ROBERTSSON94,XU95}.


\subsection{Evolution equations}\label{SecIbvpEDP}

To obtain the evolution equations satisfied by $\sigma_{ij}$, the constitutive law (\ref{SigmaIJ}) is differentiated in terms of $t$, taking (\ref{Psi2D}). If $i=j$, we obtain
\begin{equation}
\frac{\textstyle \partial\,\sigma_{ij}}{\textstyle \partial\,t}=\left(\pi_u-2\,\mu_u\right)\,\frac{\textstyle \partial\,v_k}{\textstyle \partial\,x_k}+2\,\mu_u\,\frac{\textstyle \partial\,v_i}{\textstyle \partial\,x_j}+\sum_{\ell=1}^{N_r}\xi_{ij\ell},
\label{dSdt1}
\end{equation}
where the $\xi_{ij\ell}$ are called {\it memory variables}
\begin{equation}
\xi_{ij\ell}=-\theta_\ell\,\left(\pi_r\,\kappa_\ell^p-2\,\mu_r\,\kappa_\ell^s\right)e^{-\theta_\ell\,t}H(t)*\frac{\textstyle \partial\,v_k}{\textstyle \partial\,x_k}-2\,\mu_r\,\theta_\ell\,\kappa_\ell^s\,e^{-\theta_\ell\,t}H(t)*\frac{\textstyle \partial\,v_i}{\textstyle \partial\,x_j}.
\label{Xi1}
\end{equation}
These memory variables satisfy the differential equations
\begin{equation}
\frac{\textstyle d\,\xi_{ij\ell}}{\textstyle d\,t}=-\theta_\ell\,\left(\xi_{ij\ell}+\left(\pi_r\,\kappa_\ell^p-2\,\mu_r\,\kappa_\ell^s\right)\frac{\textstyle \partial\,v_k}{\textstyle \partial\,x_k}+2\,\mu_r\,\kappa_\ell^s\,\frac{\textstyle \partial\,v_i}{\textstyle \partial\,x_j}\right),\qquad \ell=1,...,\,N_r.
\label{dKsidt1}
\end{equation}
In the same way, if $i\neq j$, we obtain
\begin{equation}
\frac{\textstyle \partial\,\sigma_{ij}}{\textstyle \partial\,t}=\mu_u\left(\frac{\textstyle \partial\,v_i}{\textstyle \partial\,x_j}+\frac{\textstyle \partial\,v_j}{\textstyle \partial\,x_i}\right)+\sum_{\ell=1}^{N_r}\xi_{ij\ell},
\label{dSdt2}
\end{equation}
with the memory variables 
\begin{equation}
\xi_{ij\ell}=-\mu_r\,\theta_\ell\,\kappa_\ell^s\,e^{-\theta_\ell\,t}H(t)*\left(\frac{\textstyle \partial\,v_i}{\textstyle \partial\,x_j}+\frac{\textstyle \partial\,v_j}{\textstyle \partial\,x_i}\right),
\label{Xi2}
\end{equation}
that satisfy the differential equations
\begin{equation}
\frac{\textstyle d\,\xi_{ij\ell}}{\textstyle d\,t}=-\theta_\ell\left(\xi_{ij\ell}+\mu_r\,\kappa_\ell^s\left(\frac{\textstyle \partial\,v_i}{\textstyle \partial\,x_j}+\frac{\textstyle \partial\,v_j}{\textstyle \partial\,x_i}\right)\right),\qquad \ell=1,...,\,N_r.
\label{dKsidt2}
\end{equation}
The convolutions in (\ref{Xi1}) and (\ref{Xi2}) induced by the convolution in (\ref{SigmaIJ}) are no longer involved in (\ref{dKsidt1}) and (\ref{dKsidt2}): adding a set of memory variables that satisfy local-in-time differential equations avoids to store the past values of the solution. In 2-D contexts, combining (\ref{dSdt1}), (\ref{dKsidt1}), (\ref{dSdt2}) and (\ref{dKsidt2}) with Newton's law yields a system of $5+3\,N_r$ partial differential equations
\begin{equation}
\left\{
\begin{array}{l}
\displaystyle 
\frac{\textstyle \partial \,v_1}{\textstyle \partial \,t}-\frac{\textstyle 1}{\textstyle \rho}\left(\frac{\textstyle \partial\,\sigma_{11}}{\textstyle \,\partial\, x} +\frac{\textstyle \partial\,\sigma_{12}}{\textstyle \,\partial\, y}\right)=0,\\
[9pt]
\displaystyle 
\frac{\textstyle \partial \,v_2}{\textstyle \partial \,t}-\frac{\textstyle 1}{\textstyle \rho}\left(\frac{\textstyle \partial\,\sigma_{12}}{\textstyle \,\partial\, x} +\frac{\textstyle \partial\,\sigma_{22}}{\textstyle \,\partial\, y}\right)=0,\\
[6pt]
\displaystyle 
\frac{\textstyle \partial \,\sigma_{11}}{\textstyle \partial \,t}-\pi_u\,\frac{\textstyle \partial \,v_1}{\textstyle \partial \,x}-(\pi_u-2\,\mu_u)\,\frac{\textstyle \partial \,v_2}{\textstyle \partial \,y}=\sum_{\ell=1}^{N_r}\xi_{11\ell} ,\\
[6pt]
\displaystyle 
\frac{\textstyle \partial \,\sigma_{12}}{\textstyle \partial \,t}-\mu_u\left(\frac{\textstyle \partial \,v_1}{\textstyle \partial \,y}+\frac{\textstyle \partial \,v_2}{\textstyle \partial \,x}  \right)=\sum_{\ell=1}^{N_r}\xi_{12\ell},\\
[6pt]
\displaystyle 
\frac{\textstyle \partial \,\sigma_{22}}{\textstyle \partial \,t}-(\pi_u-2\,\mu_u)\,\frac{\textstyle \partial \,v_1}{\textstyle \partial \,x}-\pi_u\,\frac{\textstyle \partial \,v_2}{\textstyle \partial \,y}=\sum_{\ell=1}^{N_r}\xi_{22\ell},\\
[12pt]
\displaystyle 
\frac{\textstyle \partial \,\xi_{11\ell}}{\textstyle \partial \,t}+ \theta_\ell\,\left(\pi_r\,\kappa_\ell^p \frac{\textstyle \partial\,v_1}{\textstyle \partial\,x}+\left(\pi_r\,\kappa_\ell^p-2\,\mu_r\,\kappa_\ell^s\right)\frac{\textstyle \partial\,v_2}{\textstyle \partial\,y}\right)=-\theta_\ell\,\xi_{11\ell} , \quad \ell=1,...,\,N_r\\
[12pt]
\displaystyle 
\frac{\textstyle \partial \,\xi_{12\ell}}{\textstyle \partial \,t}+\mu_r\,\theta_\ell\,\kappa_\ell^s\,\left(\frac{\textstyle \partial\,v_1}{\textstyle \partial\,y}+\frac{\textstyle \partial\,v_2}{\textstyle \partial\,x}\right)=-\theta_\ell\,\xi_{12\ell} , \quad \ell=1,...,\,N_r\\
[12pt]
\displaystyle 
\frac{\textstyle \partial \,\xi_{22\ell}}{\textstyle \partial \,t}+\theta_\ell\,\left(\left(\pi_r\,\kappa_\ell^p-2\,\mu_r\,\kappa_\ell^s\right)\frac{\textstyle \partial\,v_1}{\textstyle \partial\,x}+\pi_r\,\kappa_\ell^p\,\frac{\textstyle \partial\,v_2}{\textstyle \partial\,y}\right)=-\theta_\ell\,\xi_{22\ell} , \quad \ell=1,...,\,N_r.
\end{array}
\right.
\label{LC}
\end{equation}
Setting
\begin{equation}
{\bf U}=\left(v_1,\,v_2,\, \sigma_{11},\,\sigma_{12},\,\sigma_{22},\, \xi_{111},...,\,\xi_{11N_r},\,\xi_{121},...,\,\xi_{12N_r},\,\xi_{221},...\,\xi_{22N_r}\right)^T,
\label{Ucomplet}
\end{equation}
one can write (\ref{LC}) in the form of a first-order linear system with a source term
\begin{equation}
\frac{\textstyle \partial}{\textstyle \partial\,t}\,{\bf U}+{\bf A}\,\frac{\textstyle \partial}{\textstyle \partial\,x}\,{\bf U}+{\bf B}\,\frac{\textstyle \partial}{\textstyle \partial\,y}\,{\bf U}=-{\bf S}\,{\bf U},
\label{SystUnsplit}
\end{equation}
where ${\bf A}$, ${\bf B}$ and ${\bf S}$ are $(5+3\,N_r)\times(5+3\,N_r)$ matrices. The eigenvalues of ${\bf A}$ and ${\bf B}$ are real: $\pm c_{p\infty}$, $\pm c_{s\infty}$, and 0 with multiplicity $3\,N_r+1$. As deduced from (\ref{Theta}), the spectral radius of ${\bf S}$ is 
\begin{equation}
R({\bf S})=\theta_{N_r}=\frac{\textstyle \omega_1}{\textstyle 2\,\pi}=f_1.
\label{RayonS}
\end{equation} 
For further use, we introduce the restriction of ${\bf U}$ to the velocity and stress components and without any memory variables:
\begin{equation}
\overline{\bf U}=\left(v_1,\,v_2,\, \sigma_{11},\,\sigma_{12},\,\sigma_{22}\right)^T.
\label{Urestreint}
\end{equation}
An overline is also used to denote the restricted $5 \times 5$ matrices $\overline{\bf A}$ and $\overline{\bf B}$ involving only the velocity and stress components.

Even in non-viscoelastic subdomains, the evolution equations are written in the same way as (\ref{SystUnsplit}). For instance, $\Omega_1$ is assumed to be a fluid medium in section \ref{SecIbvpJC} and in the numerical experiments. In this case, ${\bf U}=\overline{\bf U}=(v_1,\,v_2,\,p)^T$, where $p$ is the acoustic pressure, ${\bf A}$ and ${\bf B}$ are $3 \times 3$ matrices, and ${\bf S}={\bf 0}$. Lastly, subscripts will be used to denote the medium under investigation: as an example, ${\bf A}_0$ is the matrix ${\bf A}$ in $\Omega_0$.


\subsection{Interface conditions}\label{SecIbvpJC}

\begin{figure}[htbp]
\begin{center}
\includegraphics[scale=0.7]{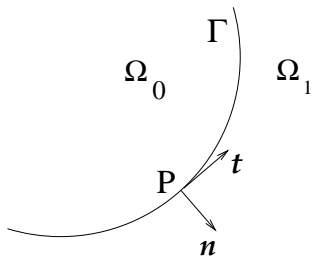} 
\caption{Interface $\Gamma$ between two media $\Omega_0$ et $\Omega_1$.}
\label{Crobar}
\end{center}
\end{figure}

The physical parameters defined in section \ref{SecIbvpEDP} can vary discontinuously across interfaces. In what follows, we will focus on two domains $\Omega_0$ and $\Omega_1$, which are separated by a stationary interface $\Gamma$ described by a parametric equation $(x(\tau),\,y(\tau))$ (figure \ref{Crobar}). The domain $\Omega_0$ contains a viscoelastic medium described by the generalized Zener model: the constitutive law and the evolution equations in $\Omega_0$ are described in sections \ref{SecPbZener} and \ref{SecIbvpEDP}. The domain $\Omega_1$ can contain vacuum, a perfect fluid, an elastic solid or any other viscoelastic medium: all these combinations have been implemented numerically and tested. In the rest of the study, we will focus on the case where $\Omega_1$ contains a fluid. In this case, the interface conditions are
\begin{equation}
[{\bf v.n}]=0,\qquad ({\bf \sigma.n}).{\bf n}=-p.{\bf n}^2,\qquad ({\bf \sigma.n}).{\bf t}=0,
\label{JCvisco/eau}
\end{equation}
where $[.]$  refers to the jump from $\Omega_0$ to $\Omega_1$, and the unit tangential vector ${\bf t}$ and the unit normal vector ${\bf n}$ are 
\begin{equation}
{\bf t}=
\frac{\textstyle 1}{\textstyle \sqrt{x^{'2}+y^{'2}}}\,
\left(
x^{'},\,\,y^{'}
\right)^T,\qquad
{\bf n}=
\frac{\textstyle 1}{\textstyle \sqrt{x^{'2}+y^{'2}}}\,
\left(
y^{'},\,-x^{'}
\right)^T.
\label{NT}
\end{equation}
Derivatives $x^{'}=\frac{d\,x}{d\,\tau}$ and $y^{'}=\frac{d\,y}{d\,\tau}$ are assumed to be continuous everywhere along $\Gamma$, and to be differentiable as many times as required. 

The interface conditions (\ref{JCvisco/eau}) only involve velocity and stress components $\overline{\bf U}$ (\ref{Urestreint}). This is also the case for other configurations, such as viscoelastic / vacuum or 
viscoelastic / viscoelastic interfaces. In section \ref{SecMethodsESIM} and \ref{SecDetailsEsim}, we will write the interface conditions satisfied by the spatial derivatives of $\overline{\bf U}$ up to the $k$-th order, and hence the following notation is introduced:
\begin{equation}
\begin{array}{l}
\displaystyle
{\bf U}_\ell^k=\lim_{M\rightarrow P,\,M\in\Omega_\ell}
\left(
\overline{\bf U}^T,
...,\,
\frac{\textstyle \partial^\alpha}{\textstyle \partial\, x^{\alpha-\beta}\,\partial\,y^\beta}\,\overline{\bf U}^T,
...,\,
\frac{\textstyle \partial^k}{\textstyle \partial\,y^k}\,\overline{\bf U}^T
\right)^T,
\label{Uk}
\end{array}
\end{equation}
where $\alpha=0,\,...,\,k$, $\beta=0,\,...,\,\alpha$, and $\ell=0,\,1$ denotes the number of the medium $\Omega_\ell$. The interface conditions are then written
\begin{equation}
\begin{array}{l}
\displaystyle
{\bf C}_1^0\,{\bf U}_1^0={\bf C}_0^0\,{\bf U}_0^0,\\
[12pt]
\displaystyle
{\bf L}_1^0\,{\bf U}_1^0={\bf 0},\qquad {\bf L}_0^0\,{\bf U}_0^0={\bf 0},
\end{array}
\label{JC0}
\end{equation}
where ${\bf C}_\ell^0$ are the matrices of the jump conditions, and ${\bf L}_\ell^0$ are the matrices of the boundary conditions. With this formalism, the viscoelastic / fluid interface conditions (\ref{JCvisco/eau}) and the vectors (\ref{NT}) yield
\begin{equation}
\begin{array}{l}
\displaystyle
{\bf C}_0^0(\tau)=
\left(
\begin{array}{ccccc}
y^{'}  & -x^{'}& 0       & 0              & 0\\
[8pt]
0      & 0     & y^{'^2} & -2\,x^{'}y^{'} & x^{'^2}
\end{array}
\right),\\
\\
\displaystyle
{\bf C}_1^0(\tau)=
\left(
\begin{array}{ccccc}
y^{'} & -x^{'} & 0 \\
[8pt]
0      & 0     & -\left(x^{'2}+y^{'2}\right)
\end{array}
\right),\\
\\
\displaystyle
{\bf L}_0^0(\tau)=
\left(
\begin{array}{ccccc}
0 & 0 & x^{'}y^{'} & y^{'2}-x^{'^2} & -x^{'}y^{'}
\end{array}
\right),\quad 
{\bf L}_1^0(\tau)=
\left(
\begin{array}{ccccc}
0 & 0 & 0
\end{array}
\right).
\end{array}
\end{equation}
To conclude on that topic, it is noticed that no interface conditions are imposed on memory variables. In some cases, however, authors exhibit boundary conditions satisfied by $\xi_{ij\ell}$: see, for instance, equation (4) of \cite{ROBERTSSON96} in the case of viscoelastic / vacuum interface. Such conditions are not required to get a well-posed problem. Moreover, they are deduced from the original boundary conditions satisfied by ${\bf \sigma}$, the constitutive law (\ref{SigmaIJ}), the positivity of relaxation functions, and the definition of memory variables (\ref{Xi1}) and (\ref{Xi2}). These additional boundary conditions are not useful in the immersed interface method (section \ref{SecMethodsESIM}).


\section{Numerical modeling}\label{SecMethods}

\subsection{Numerical scheme}\label{SecMethodsADER}

{\bf Generalities}. Let us take a uniform grid, with the spatial mesh size $\Delta\,x=\Delta\,y$ and the time step $\Delta\,t$. An approximation ${\bf U}_{i,j}^n$ of ${\bf U}(x_i=i\,\Delta\,x,\,y_j=j\,\Delta\,y,\,t_n=n\,\Delta\,t)$ is sought. The numerical methods recalled in section \ref{SecIntro} usually consist in simultaneously discretizing the propagating part and the source term in (\ref{SystUnsplit}). This approach has two drawbacks. First, building unsplit methods for (\ref{LC}) is a difficult task \cite{LEVEQUE02}, whereas large classes of methods already exist for hyperbolic systems without the source term ${\bf S\,U}$. Secondly, a Von-Neumann stability analysis typically yields
\begin{equation}
\Delta\,t\leq\min\left(\frac{\textstyle \gamma\,\Delta\,x}{\textstyle c_{p\infty}},\,\frac{\textstyle 2}{\textstyle R({\bf S})}\right),
\label{CFLdirect}
\end{equation}
where $\gamma$ depends on the scheme. Based on (\ref{RayonS}) and (\ref{CFLdirect}), the spectral radius of ${\bf S}$ induces a more restrictive bound than the classical CFL condition if $f_1\geq 2\,c_{p_\infty}\,/\,(\gamma\,\Delta\,x)$, where $f_1$ is the maximum frequency considered during the determination of the parameters (section \ref{SecPbEstime}). The efficiency of the scheme is therefore penalized if large values of $f_1$ are taken.

{\bf Splitting}. Here we choose another strategy based on solving alternatively 
\begin{equation}
\left\{
\begin{array}{l}
\displaystyle
\frac{\textstyle \partial}{\textstyle \partial\,t}\,{\bf U}+{\bf A}\,\frac{\textstyle \partial}{\textstyle \partial\,x}\,{\bf U}+{\bf B}\,\frac{\textstyle \partial}{\textstyle \partial\,y}\,{\bf U}={\bf 0}, \hspace{0.5cm} (a)\\
[10pt]
\displaystyle
\frac{\textstyle \partial}{\textstyle \partial\,t}\,{\bf U}=-{\bf S}\,{\bf U}. \hspace{3.8cm} (b)
\end{array}
\right.
\label{SplittingLC}
\end{equation}
The discrete operators used in stages $(a)$ and $(b)$ are denoted by ${\bf H}_{a}$ and ${\bf H}_{b}$, respectively. The algorithm of ${\cal N}$-th order splitting is written 
\begin{equation}
\begin{array}{lllll}
\displaystyle
&\bullet&
{\bf U}_{i,j}^{(0)}&=&{\bf U}_{i,j}^n,\\
[6pt]
\displaystyle
&\bullet& {\bf U}_{i,j}^{(2m-1)}&=&{\bf H}_{a}(c_m\,\Delta\,t)\,{\bf U}_{i,j}^{(2m-2)},\qquad m=1,\cdots,\,{\cal N}\\
[6pt]
\displaystyle
&&{\bf U}_{i,j}^{(2m)}&=&{\bf H}_{b}(d_m\,\Delta\,t)\,{\bf U}_{i,j}^{(2m-1)},\\
[6pt]
\displaystyle
&\bullet& {\bf U}_{i,j}^{n+1}&=&{\bf U}_{i,j}^{(2{\cal N})},
\end{array}
\label{AlgoSplitting1}
\end{equation}
or equivalently
\begin{equation}
\displaystyle
{\bf U}_{i,j}^{n+1}=\left(\prod_{m=1}^{\cal N}{\bf H}_{b}\left(d_{{\cal N}-m+1}\,\Delta\,t\right)\,\circ\,{\bf H}_{a}\left(c_{{\cal N}-m+1}\,\Delta\,t\right)\right)\,{\bf U}_{i,j}^n.
\label{AlgoSplitting2}
\end{equation}
The coefficients $c_m$ and $d_m$ in (\ref{AlgoSplitting1}) and (\ref{AlgoSplitting2}) are given in   \ref{SecCoeffSplitting}. The cases ${\cal N}=2$ and ${\cal N}=4$ are called {\it Strang splitting} and {\it Ruth splitting}, respectively.\\

{\bf Solvers}. To solve the propagative stage (\ref{SplittingLC})-$(a)$, many standard solvers for hyperbolic systems can be used as a discrete operator ${\bf H}_{a}$. Here we choose a fourth-order ADER scheme \cite{SCHWARTZKOPFF04}. This is an explicit two time step spatially-centered flux-conserving scheme, with a centered stencil of 25 nodes. On Cartesian grids, this scheme amounts to a fourth-order Lax-Wendroff scheme. It is dispersive of order 4 and dissipative of order 6, with a stability limit $\gamma=\max(c_{p_\infty}\,\Delta\,t\,/\,\Delta\,x)=1$ \cite{HDR-LOMBARD}. 

The diffusive stage (\ref{SplittingLC})-$(b)$ is solved exactly. Based on the notations of (\ref{AlgoSplitting1}), we obtain 
\begin{equation}
\begin{array}{l}
\displaystyle
v_p^{(2m)}=v_p^{(2m-1)},\\
[5pt]
\displaystyle
\sigma_{pq}^{(2m)}=\sigma_{pq}^{(2m-1)}+\sum_{\ell=1}^{N_r}\frac{\textstyle 1}{\textstyle \theta_\ell}\left(1-e^{-\theta_\ell\,d_m\,\Delta\,t}\right)\,\xi_{pq\ell}^{(2m-1)},\\
[15pt]
\displaystyle
\xi_{pq\ell}^{(2m)}=e^{-\theta_l\,d_m\,\Delta\,t}\,\xi_{pq\ell}^{(2m-1)},
\end{array}
\label{ODEexact}
\end{equation}
where $(p,\,q)=\left\{(1,1),\,(1,2),\,(2,2)\right\}$, $\ell=1,...,\,N_r$ and $m=1,\cdots,\,{\cal N}$. The computation of $e^{-\theta_\ell\,d_m\,\Delta\,t}$ is time-consuming. To make the computational time of the diffusive stage $(b)$ negligible, the exponentials in (\ref{ODEexact}) are computed and stored once in each viscoelastic subdomain and at each time substep.  

{\bf Choice of ${\cal N}$}. The splitting (\ref{SplittingLC}) with operators ${\bf H}_{a}$ and ${\bf H}_{b}$ is ${\cal N}$-th order accurate. If ${\cal N}=1$ or ${\cal N}=2$, the optimal CFL condition of ADER scheme is recovered:$\gamma=1$. If ${\cal N}=3$ or ${\cal N}=4$, then the stability limit is improved: numerical experiments indicate $\gamma \approx 1.54$ and $\gamma \approx 1.60$, respectively. To choose ${\cal N}$, the following remarks are done:
\begin{enumerate}
\item for almost the same computational cost, ${\cal N}=2$ is twice accurate than ${\cal N}=1$, and hence Strang splitting is preferred to first-order splitting;
\item ${\cal N}=4$ maintains the fourth-order accuracy of ADER scheme. However, four spatial integrations are required, compared with a single one when ${\cal N}=2$;
\item in counterpart, coarser spatial and temporal grids can be used with Ruth splitting to ensure the same accuracy than with Strang splitting: the CPU time and the number of time steps are reduced accordingly. Moreover, the CFL limit of stability is improved with Ruth splitting.
\end{enumerate}
Further discussions on this topic and convergence measurements will be proposed in section \ref{SecNum1D}.


\subsection{Immersed interface method}\label{SecMethodsESIM}

To solve the propagative part (\ref{SplittingLC})-$(a)$ accurately, the solution must be sufficiently smooth on the whole stencil. At the irregular points, where the stencil crosses the interface, the smoothness requirement is not satisfied, and the discrete operator ${\bf H}_a$ (\ref{AlgoSplitting1}) needs to be modified to maintain the accuracy. For that purpose, an immersed interface method is implemented \cite{PIRAUX01,LOMBARD04,LOMBARD06}. 

\begin{figure}[htbp]
\begin{center}
\includegraphics[scale=0.6]{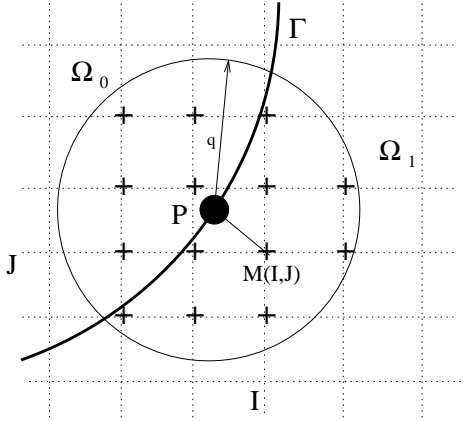}
\caption{$M(x_I,\,y_J)\in \Omega_1$ is a grid node where a modified value ${\bf U}_{I,J}^*$ is computed; $P$ is the orthogonal projection of $M$ onto the interface $\Gamma$. The grid nodes used to compute ${\bf U}_{I,J}^*$ are inside the circle with radius $q$ and centered on $P$; they are denoted by ${\bf +}$.}
\label{Patate}
\end{center}
\end{figure}

The basic principle of this method is as follows. Let us take an irregular point $(x_i,\,y_j)\in\Omega_0$. The numerical computation of ${\bf U}_{i,j}^{(2m-1)}$ in (\ref{AlgoSplitting1}) requires to use a value at $(x_I,\,y_J)\in\Omega_1$ (figure \ref{Patate}). Instead of using ${\bf U}_{I,J}^{(2m-2)}$, a modified value ${\bf U}_{I,J}^*$ is injected into the discrete operator ${\bf H}_a$. It amounts to a $k$-th order extension of the solution from $\Omega_0$ into $\Omega_1$, where $k$ is an integer to be defined.

In a few words, ${\bf U}_{I,J}^*$ is build as follows. Let $P$ be the orthogonal projection of $(x_I,\,y_J)$ on $\Gamma$, and consider the disc ${\cal D}$ centered on $P$ with a radius $q$ (figure \ref{Patate}). Based on the interface conditions (\ref{JC0}) at $P$ and on the numerical values $\overline{\bf U}_{I,J}^{(2m-2)}$ at grid nodes inside ${\cal D}$, a matrix ${\cal M}$ is defined so that
\begin{equation}
{\bf U}_{I,J}^*={\cal M}\left(\overline{\bf U}^{(2m-2)}\right)_{\mathcal D}.
\label{UIJ*}
\end{equation}
It is noticed that restrictions (\ref{Urestreint}) to velocity and stress components $\overline{\bf U}_{I,J}^{(2m-2)}$ are considered in (\ref{UIJ*}): the system (\ref{LC}) implies that no spatial derivatives are applied on the memory variables, and hence the numerical integration of (\ref{SplittingLC})-$(a)$ at $(x_i,\,y_j)$ does not involve values of the memory variables at other nodes. 

During the propagative part (\ref{SplittingLC})-$(a)$, the viscoelastic medium behaves like an elastic medium. The derivation of the matrix ${\cal M}$ in (\ref{UIJ*}) is therefore in line with the algorithm developed for non-dissipative media \cite{LOMBARD04}, with appropriate physical parameters. Details are given in \ref{SecDetailsEsim}. The only slight modification compared with the elastic case concerns the Beltrami-Michell equations: see step 2 of \ref{SecDetailsEsim}. 

Some comments are done about the immersed interface method:
\begin{enumerate}
\item A similar algorithm is applied at each irregular point along $\Gamma$ and at each propagative part of the splitting algorithm (\ref{AlgoSplitting1}) ($m=1,\cdots,\,{\cal N}$). Since the jump conditions do not vary with time, the evaluation of the matrices in (\ref{UIJ*}) is done during a preprocessing step. Only small matrix-vector products are therefore required at each splitting step. After optimization of the computer codes, this additional cost is made negligible, lower than 1\% of the time-marching.

\item The matrix ${\cal M}$ in (\ref{UIJ*}) depends on the subcell position of $P$ inside the mesh and on the jump conditions at $P$, involving the local geometry and the curvature of $\Gamma$ at $P$. Consequently, all these insights are incorporated in the modified value (\ref{UIJ*}), and hence in the scheme.

\item \label{EstimeVarEps} The simulations indicate that the number of grid nodes inside the disc ${\cal D}$ has a crucial influence on the stability of the immersed interface method. Here we use a constant radius $q$. Taking $k=2$, numerical experiments have shown that $q=3.2\,\Delta\,x$ is a good candidate, while $q=4.5\,\Delta\,x$ is used when $k=3$. 

\item The order $k$ plays an important role on the accuracy of the coupling between the immersed interface method and a $r$-th order scheme. If $k\geq r$, then a $r$-th order local truncation error is obtained at the irregular points. However, $k=r-1$ suffices to keep the global error to the $r$-th order \cite{GUSTAFSSON75}, and hence $k=3$ is used for the ADER 4 scheme. 
\end{enumerate} 


\section{Numerical experiments}\label{SecNum}

\subsection{Configuration}\label{SecNumConfig}

Here we focus on viscoelastic / fluid configurations. The physical parameters in the viscoelastic medium $\Omega_0$ are:
\begin{equation}
\rho=1200\,\mbox{ kg/m}^3,\, c_{p_0}=2800\,\mbox{ m/s},\, c_{s_0}=1400\,\mbox{ m/s},\,
Q^p_{ref}=20,\, Q^s_{ref}=15,
\label{ParamVisco2D}
\end{equation}
with $N_r=3$ relaxation mechanisms, and in the fluid medium $\Omega_1$ they are:
\begin{equation}
\rho=1000\,\mbox{ kg/m}^3,\, c=1500\,\mbox{ m/s}.
\label{ParamFluide}
\end{equation}
The time evolution of the source is given by a combination of truncated sinusoids
\begin{equation}
h(t)=
\left\{
\begin{array}{l}
\displaystyle
\displaystyle \sum_{m=1}^4 a_m\,\sin(\beta_m\,\omega_c\,t)\quad \mbox{ if  }\, 0<t<\frac{\textstyle 1}{\textstyle f_c},\\
\\
0 \,\mbox{ otherwise}, 
\end{array}
\right.
\label{JKPS}
\end{equation}
where $\beta_m=2^{m-1}$, $\omega_c=2\pi\,f_c$; the coefficients $a_m$ are: $a_1=1$, $a_2=-21/32$, $a_3=63/768$, $a_4=-1/512$. This source is $C^6$ and has a central frequency $f_c=40$ Hz. Once  propagation has occured across a viscoelastic medium, waves emitted by the source are deformed compared with (\ref{JKPS}). After optimizing $\kappa_\ell$ between $f_0=f_c/10$ and $f_1=10\,f_c$ (\ref{Emmerich}), then (\ref{Cpinf}) yields the high-frequency limits $c_{p_\infty}=3161$ m/s and $c_{s_\infty}=1645$ m/s.

The computations are performed on $N_x \times N_y$ grid nodes. The discretization mesh is $\Delta\,x=\Delta\,y=1$ m, and Strang splitting is used. The time step follows from $c_{p_\infty}\,\Delta\,t/\Delta\,x=0.85$. A third-order immersed interface method is used: $k=3$ and $q=3.2\,\Delta\,x$ in section \ref{SecMethodsESIM}. Time-domain exact solutions are not available in dissipative media; they are therefore computed by Fourier synthesis on $N_f$ modes, with a frequency step $\Delta\,f$.

On the plates, $-p$ and $\sigma_{11}$ are shown in fluid and viscoelastic media, respectively. P waves and S waves are displayed with a green-red palette and a yellow-magenta palette, respectively. The distinction between these waves is based on numerical estimates of div ${\bf v}$ and curl ${\bf v}$. The position of a slice is denoted on the plates by horizontal segments.


\subsection{Test 1: 1D medium}\label{SecNum1D}

\begin{figure}[htbp]
\begin{center}
\begin{tabular}{cc}
(a) & (b)\\
\includegraphics[scale=0.31]{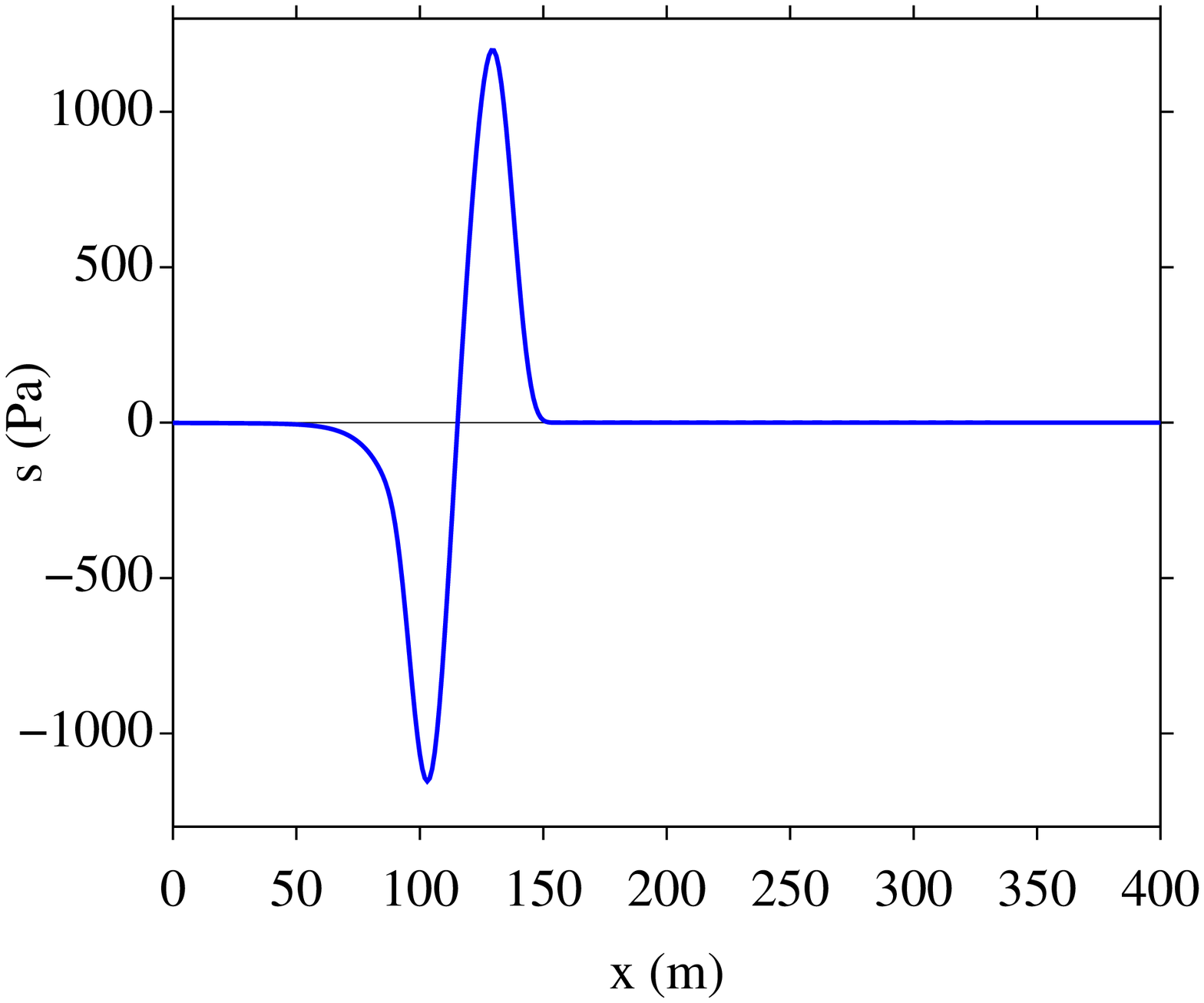}&
\includegraphics[scale=0.31]{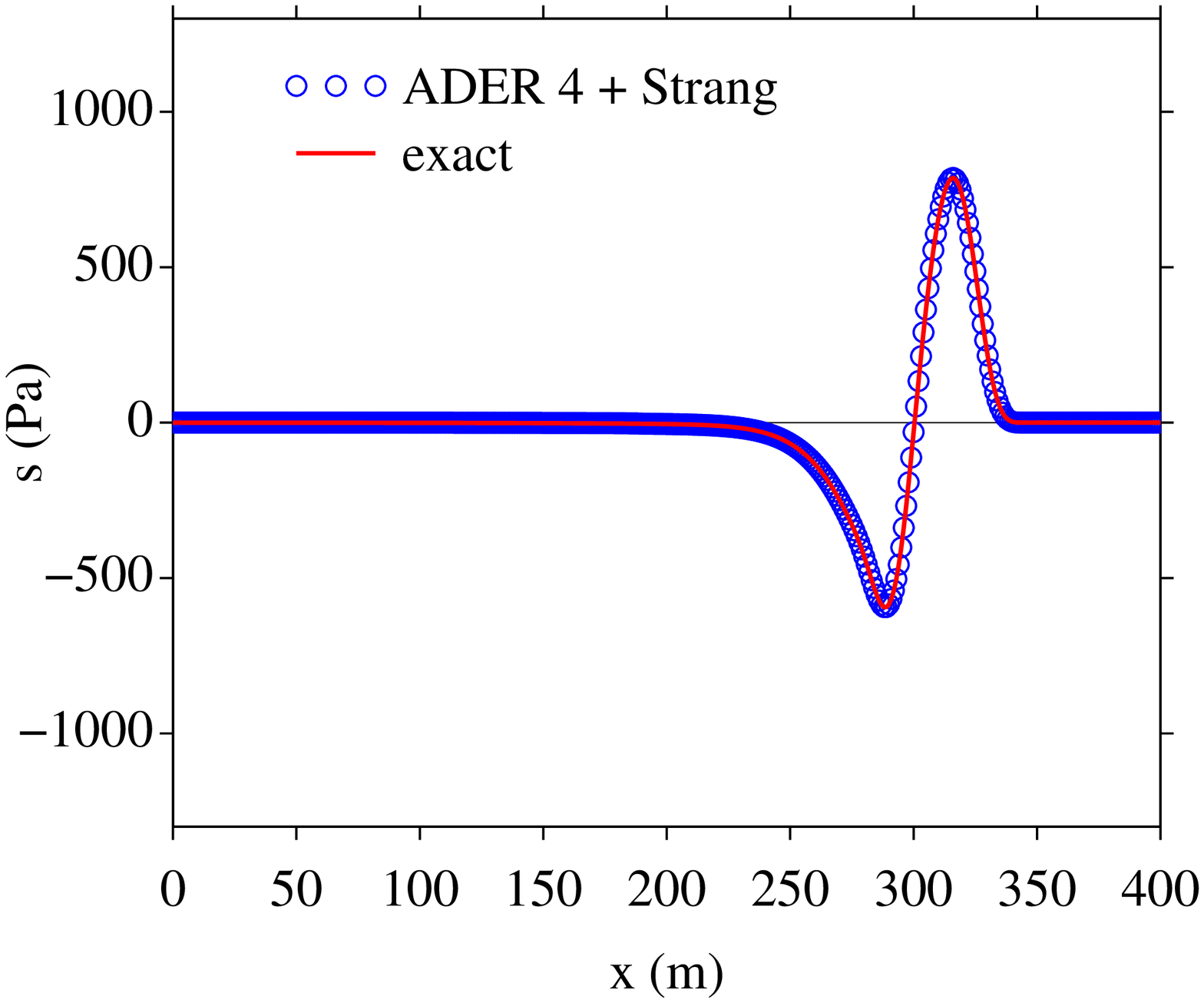}
\end{tabular}
\end{center}
\caption{test 1, homogeneous viscoelastic medium. Stress $\sigma$ at the initial instant (a) and after 200 time steps (b). }
\label{FigVisco1D}
\end{figure}

As a first experiment, wave propagation is simulated in a one-dimensional homogeneous viscoelastic medium $[0,\,400]$ m. The physical parameters are those of P-waves in (\ref{ParamVisco2D}). The initial data is the right-going part of the field emitted by a source point at $x=0$ and propagated during 0.05 s (figure \ref{FigVisco1D}-a). This field is computed by Fourier synthesis.

Figure \ref{FigVisco1D}-b shows the stress obtained by taking $N_x=400$ grid nodes and $N_t=200$ time steps. One clearly observes the attenuation and the dispersion induced by the viscoelastic constitutive law. Excellent agreement is observed between numerical and exact values. 

\begin{figure}[htbp]
\begin{center}
\begin{tabular}{c}
\includegraphics[scale=0.33]{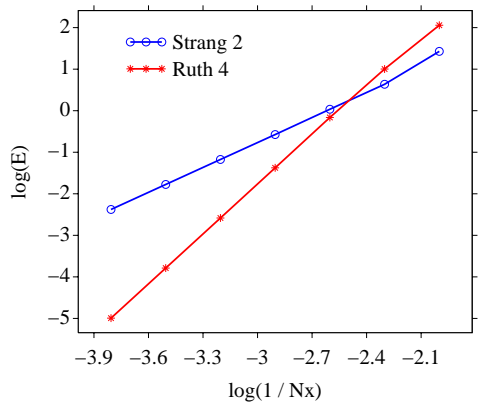}
\end{tabular}
\end{center}
\vspace{-1cm}
\caption{test 1, homogeneous viscoelastic medium. Error versus the number of grid nodes. Slopes + 2 and +4 are obtained with Strang splitting and Ruth splitting, respectively.}
\label{FigConverge1D}
\end{figure}

To study more quantitatively the accuracy of the numerical scheme, convergence measurements are performed by considering two different splittings: Strang splitting (${\cal N}=2$) and Ruth splitting (${\cal N}=4$); see section \ref{SecMethodsADER} for details. Special care is taken to ensure that the initial data and the exact solution are valuable reference solutions: $N_f=65536$ Fourier modes are used, with a frequency step $\Delta\,f=0.01$ Hz. Errors in norm $l_2$ and convergence rates are reported in table \ref{TabConverge1D}, and then are displayed in figure \ref{FigConverge1D}. For coarse grids ($N_x\leq 400$), theoretical orders are not yet reached. Moreover, Strang splitting is more accurate than Ruth splitting. For $N_x\geq 400$, second-order and fourth-order rates are very closely reached, and Ruth splitting becomes competitive.

\begin{table}[h]
\begin{center}
\begin{small}
\begin{tabular}{l|ll|ll}
$N_x$   &  ${\cal N}=2$          & rate          & ${\cal N}=4$        & rate\\
\hline
 100    & $2.680\,10^{ 1}$       & {\bf -}       & $1.141\,10^{2}$     & {\bf -}      \\
 200    & $4.325\,10^{ 0}$       & {\bf 2.631}   & $1.007\,10^{1}$     & {\bf 3.502}  \\
 400    & $1.083\,10^{ 0}$       & {\bf 1.998}   & $6.870\,10^{-1}$    & {\bf 3.874}  \\
 800    & $2.665\,10^{-1}$       & {\bf 2.023}   & $4.217\,10^{-2}$    & {\bf 4.026}  \\
1600    & $6.700\,10^{-2}$       & {\bf 1.992}   & $2.613\,10^{-3}$    & {\bf 4.012}  \\
3200    & $1.677\,10^{-2}$       & {\bf 1.998}   & $1.629\,10^{-4}$    & {\bf 4.004}  \\
6400    & $4.195\,10^{-3}$       & {\bf 1.999}   & $1.016\,10^{-5}$    & {\bf 4.003} 
\end{tabular}
\end{small} 
\end{center}
\vspace{-0.5cm}
\caption{test 1, homogeneous viscoelastic medium. Convergence measurements.}
\label{TabConverge1D}
\end{table}

\begin{figure}[htbp]
\begin{center}
\begin{tabular}{cc}
(a) & (b)\\
\includegraphics[scale=0.31]{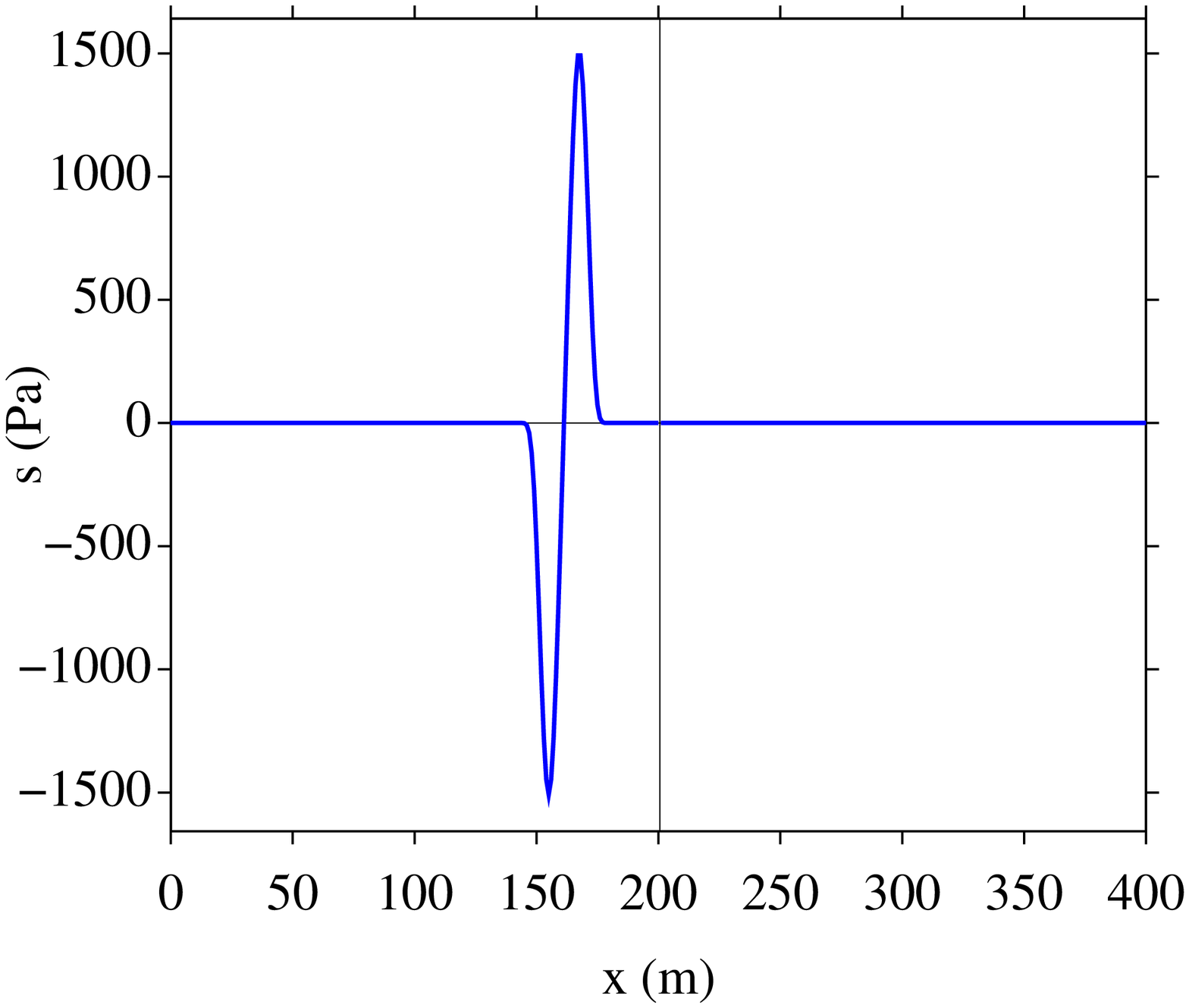}&
\includegraphics[scale=0.31]{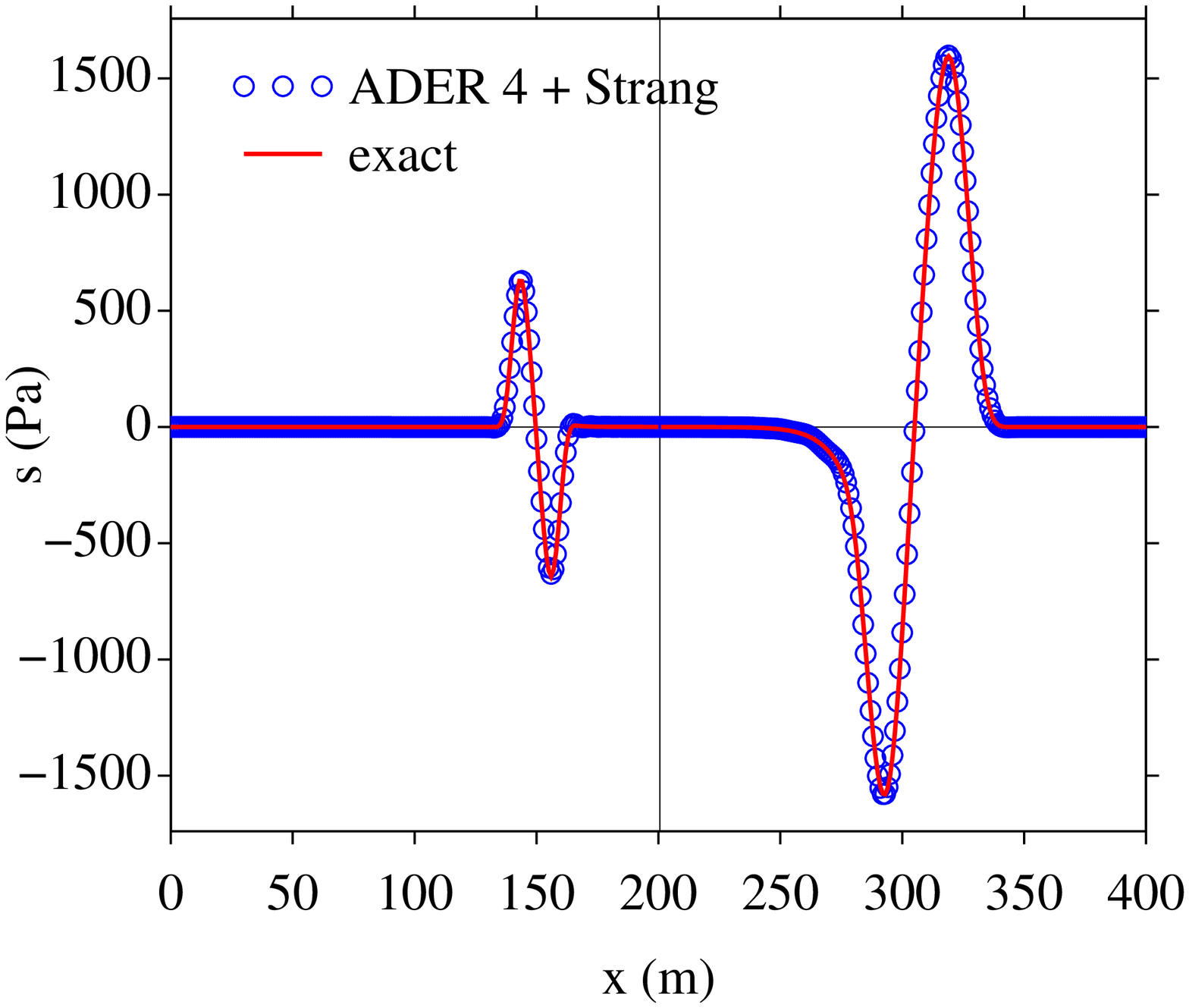}
\end{tabular}
\end{center}
\vspace{-1cm}
\caption{test 1, fluid / viscoelastic medium. Stress $\sigma$ at the initial instant (a) and after 200 time steps (b). }
\label{FigFluideVisco1D}
\end{figure}

The case of a 1D heterogeneous fluid / viscoelastic medium is illustrated in figure \ref{FigFluideVisco1D}. The interface is located at $x=200$ m. The initial right-going wave is put in the fluid medium (figure \ref{FigFluideVisco1D}-a). Wave propagation is simulated during 200 time steps. At the final instant, the incident wave has interacted with the interface, and reflected and transmitted waves have been generated. Numerical and exact solutions are compared successfully (figure \ref{FigFluideVisco1D}-b).

Convergence measurements have been performed in this heterogeneous case, but the results are not so sharp than in table \ref{TabConverge1D} and figure \ref{FigConverge1D}. When the dissipation is small $(Q \geq 100$), the results are similar to those obtained in the homogeneous case: second-order and fourth-order rates are obtained by Strang splitting and Ruth splitting, respectively. But when the dissipation becomes important, convergence rates between 1 and 2 are obtained, whatever the splitting. The problem probably follows from the coupling between the splitting and the immersed interface method, during the time derivatives of interface conditions (step 1/4 of \ref{SecDetailsEsim}). These derivatives are based on the conservation law (\ref{SplittingLC})-(a), corresponding to a relaxed elastic medium, instead of the original system (\ref{LC}). Further analysis on that topic is required, in order to propose ${\cal N}$-th order algorithms in heterogeneous cases.

\subsection{Test 2: plane wave on a plane interface}\label{SecNumPlane}

\begin{figure}[htbp]
\begin{center}
\begin{tabular}{cc}
(a) & (b)\\
\includegraphics[scale=0.33]{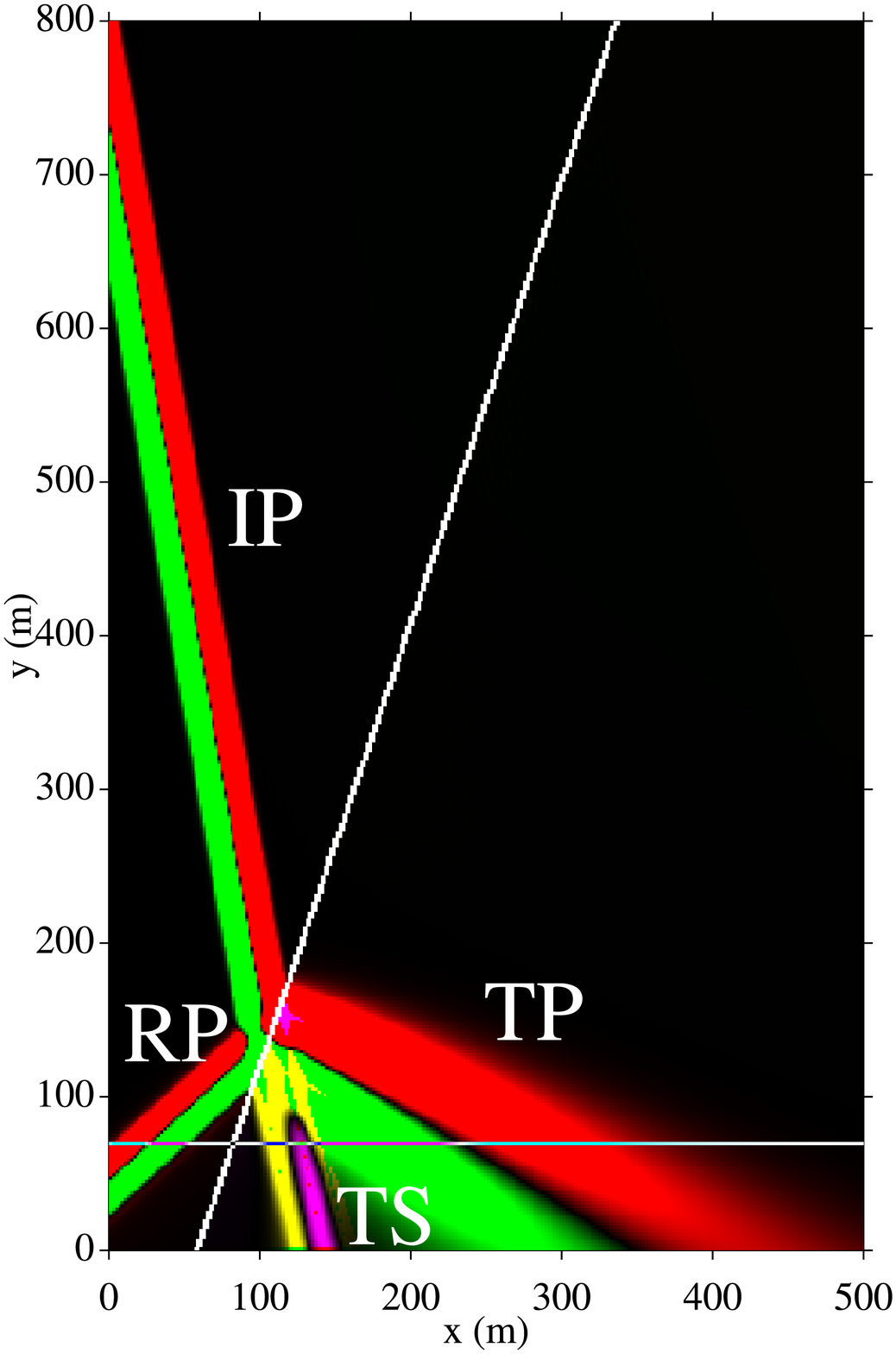}&
\includegraphics[scale=0.33]{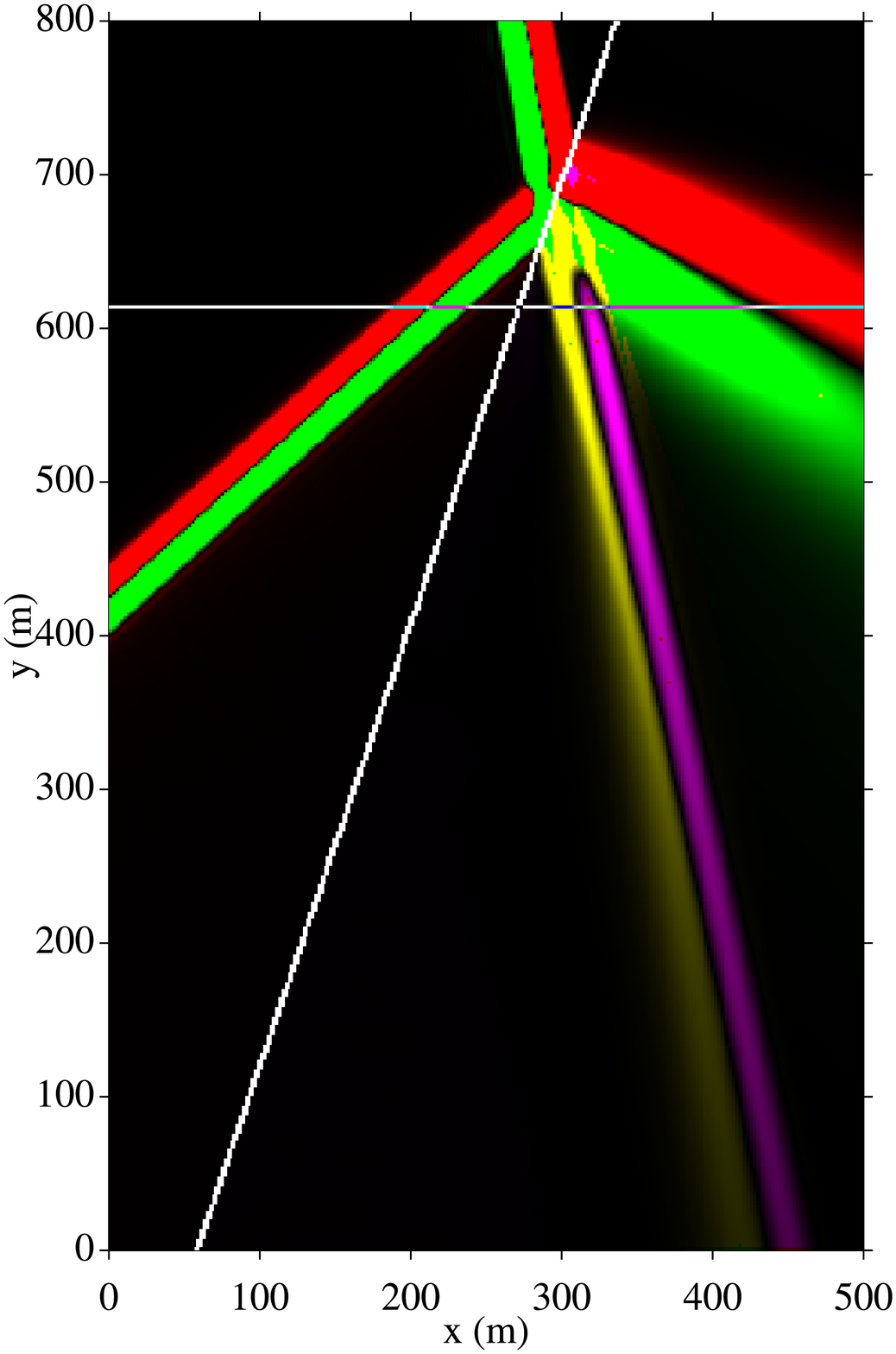}\\
(c) & (d)\\
\includegraphics[scale=0.33]{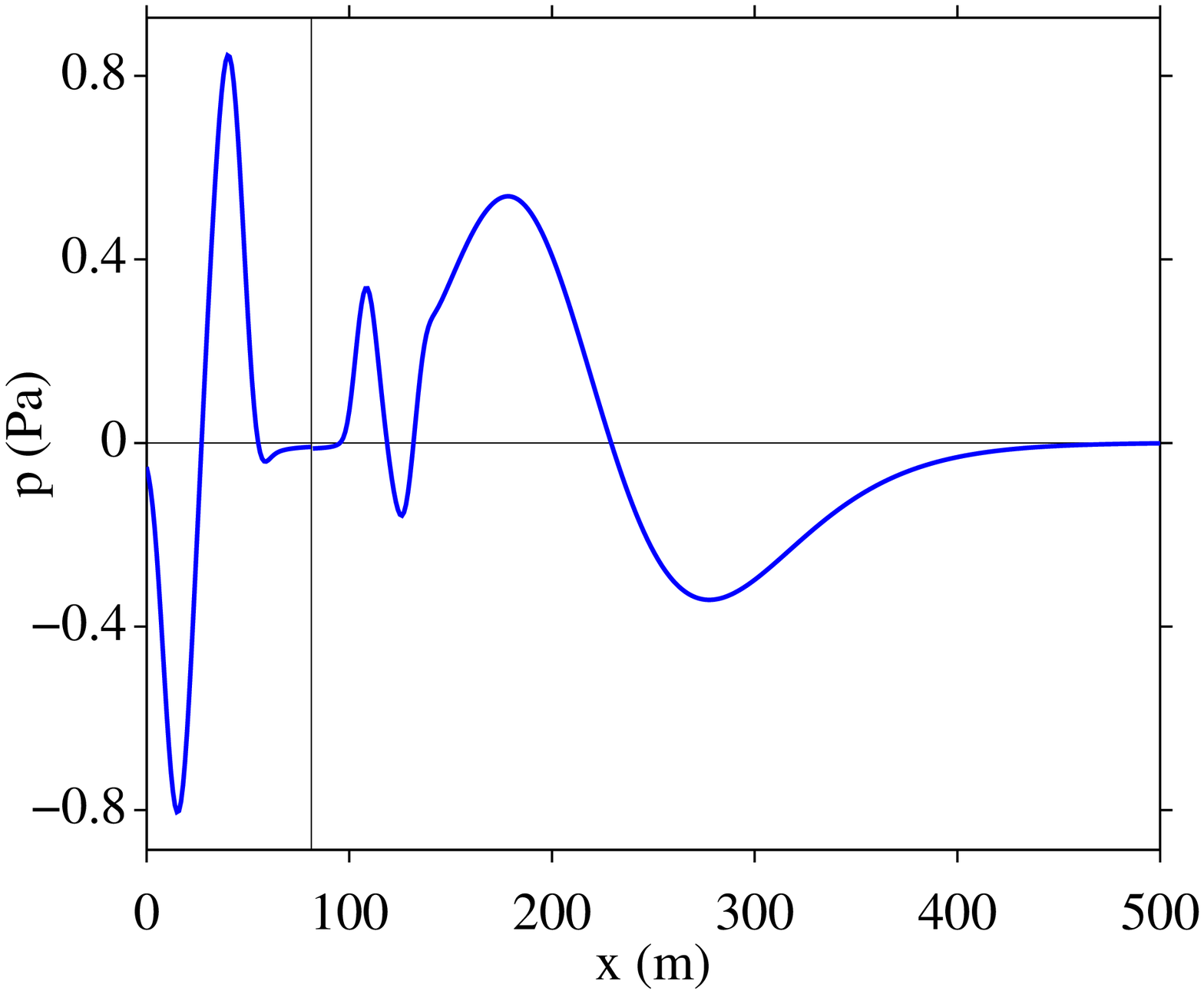}&
\includegraphics[scale=0.33]{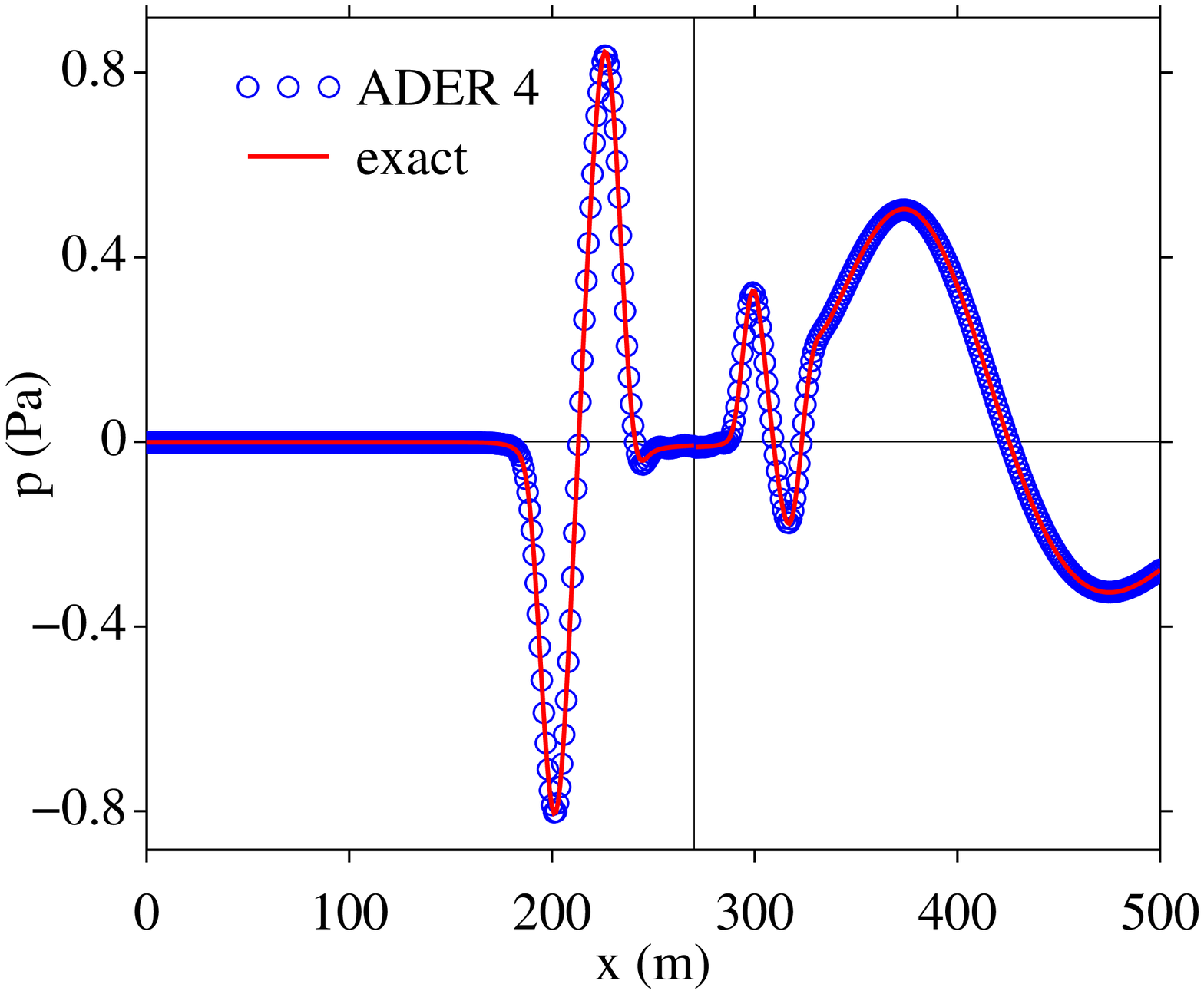}
\end{tabular}
\end{center}
\caption{test 2. Plane interface between a fluid (on the left) and a viscoelastic medium (on the right). Initial instant (a-c) and after 700 time steps (b-d). IP: incident homogeneous acoustic wave; RP: reflected inhomogeneous acoustic wave; TP,  TS: transmitted inhomogeneous compressional and shear viscoelastic waves. }
\label{FigTest1}
\end{figure}

The second test is conducted on a 2D plane interface between the fluid and the viscoelastic medium. The angle between the straight line and the horizontal axis is equal to 70 degrees. A homogeneous acoustic plane wave (IP), having a wave vector inclined at an angle of 10 degrees, propagates in the fluid and interacts with the interface. The viscoelastic transmitted compressional (TP) and shear (TS) waves are inhomogeneous waves, whose wave vector forms a non-null angle with the direction of attenuation. See \cite{LOCKETT62,COOPER66,COOPER67,BORCHERDT77,BORCHERDT82,WENNERBERG85} for further details on this topic. Figure \ref{FigTest1} shows the incident field (a-c) and after 700 integration time steps (b-d), which corresponds to roughly 6 propagation wavelengths. Excellent agreement is observed between the numerical and the exact values.


\subsection{Test 3: plane wave on a circular interface}\label{SecNumCircular}

\begin{figure}[htbp]
\begin{center}
\begin{tabular}{cc}
(a) & (b)\\
\includegraphics[scale=0.33]{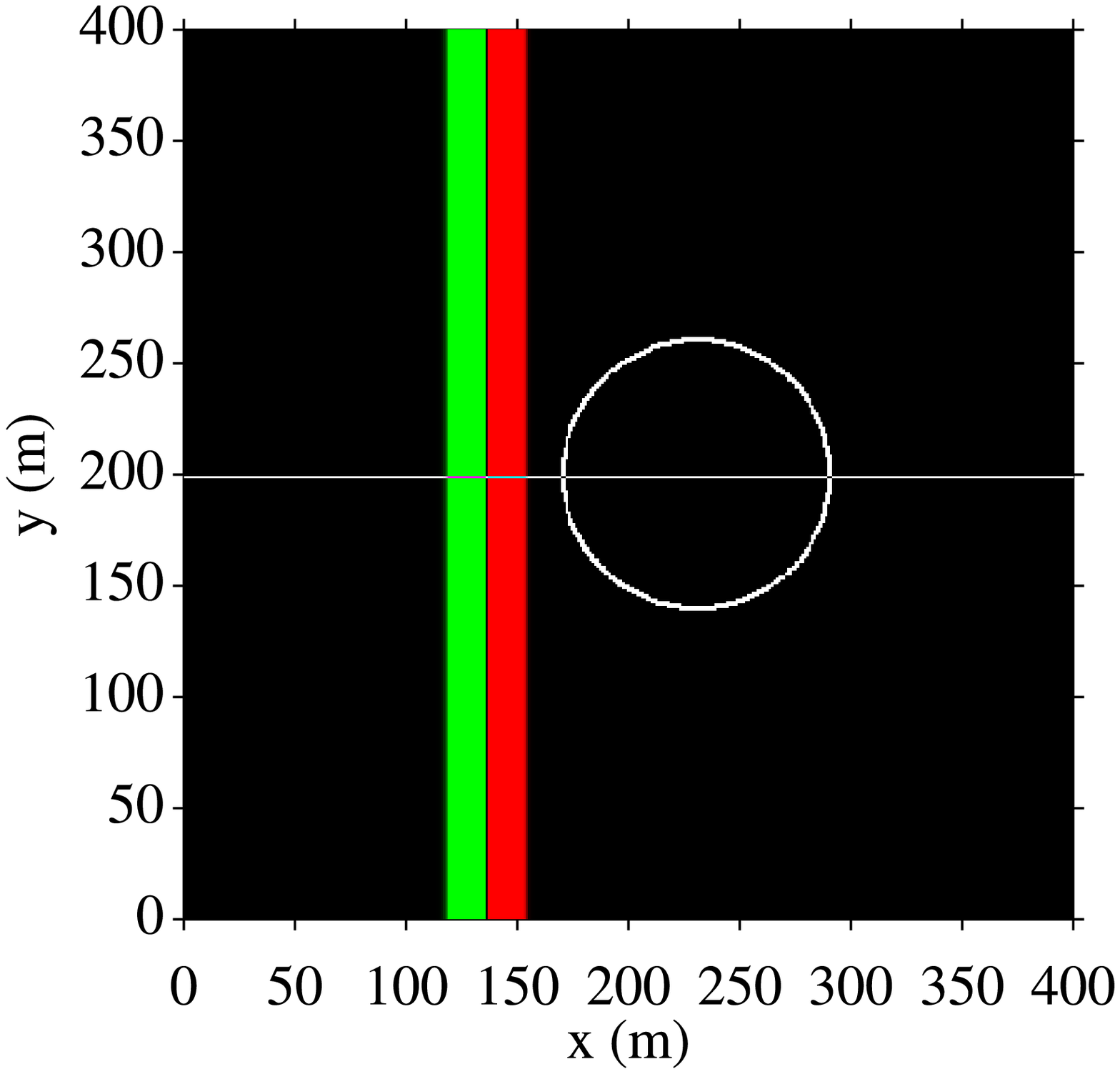}&
\includegraphics[scale=0.33]{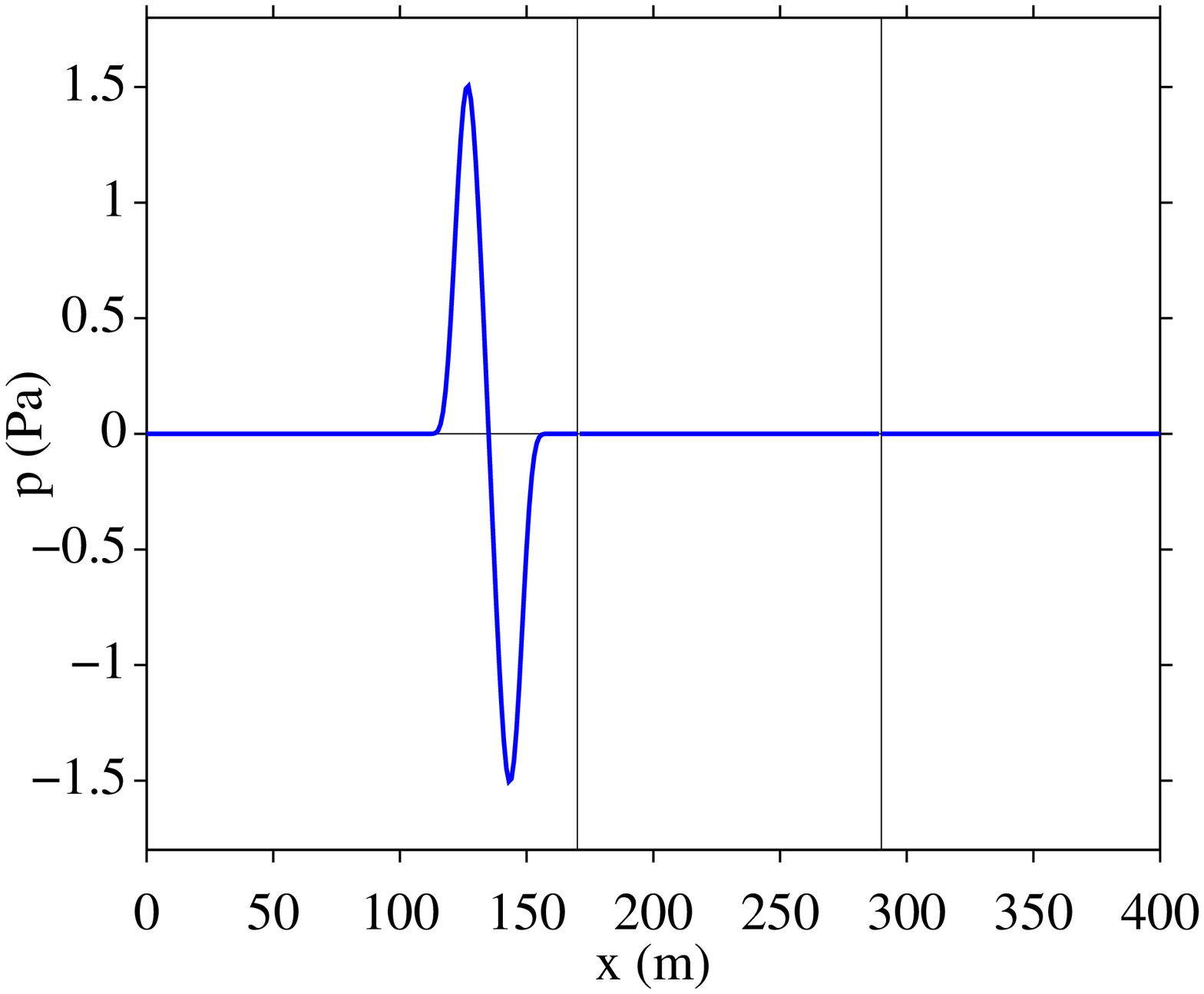}\\
(c) & (d)\\
\includegraphics[scale=0.33]{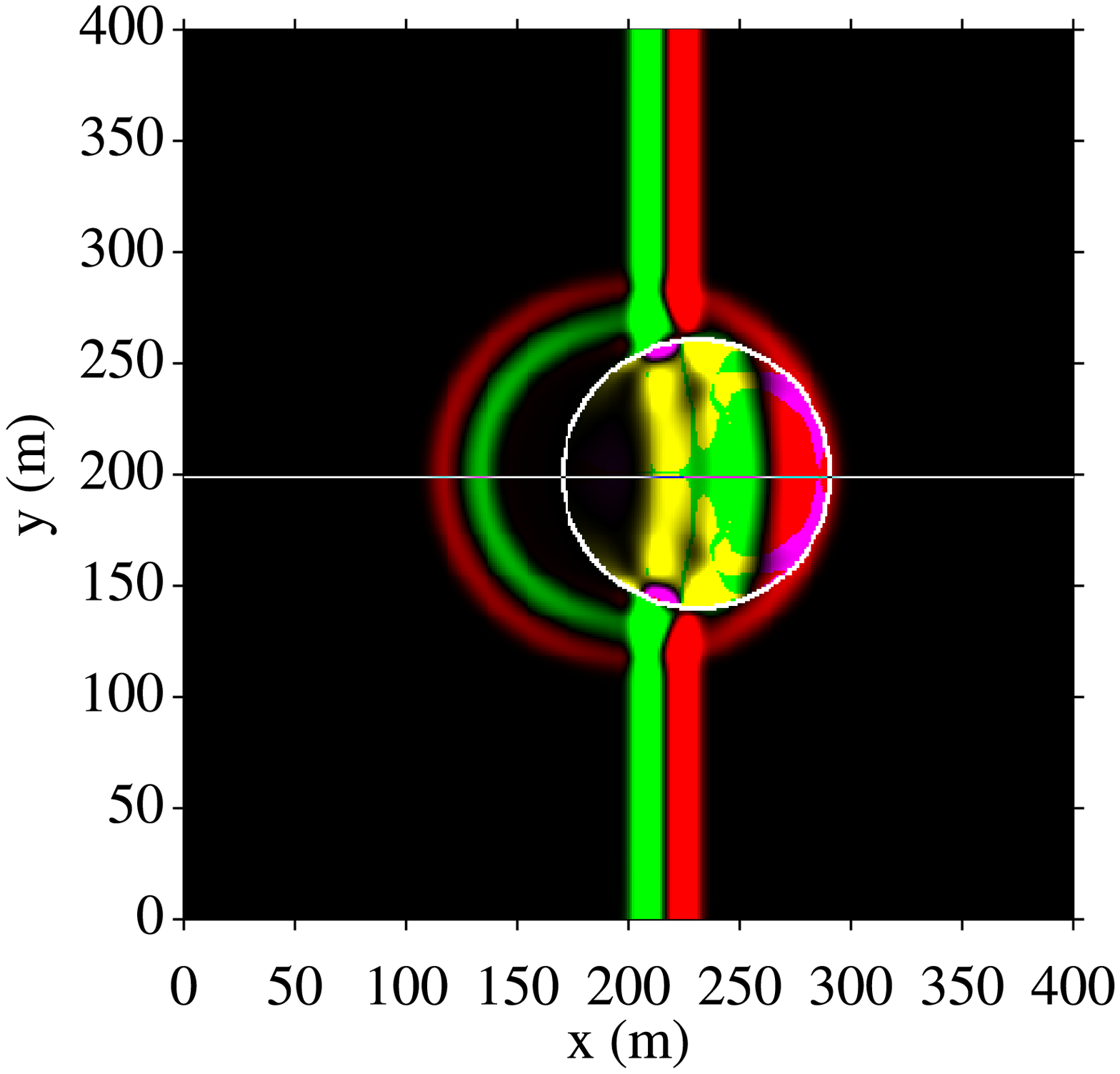}&
\includegraphics[scale=0.33]{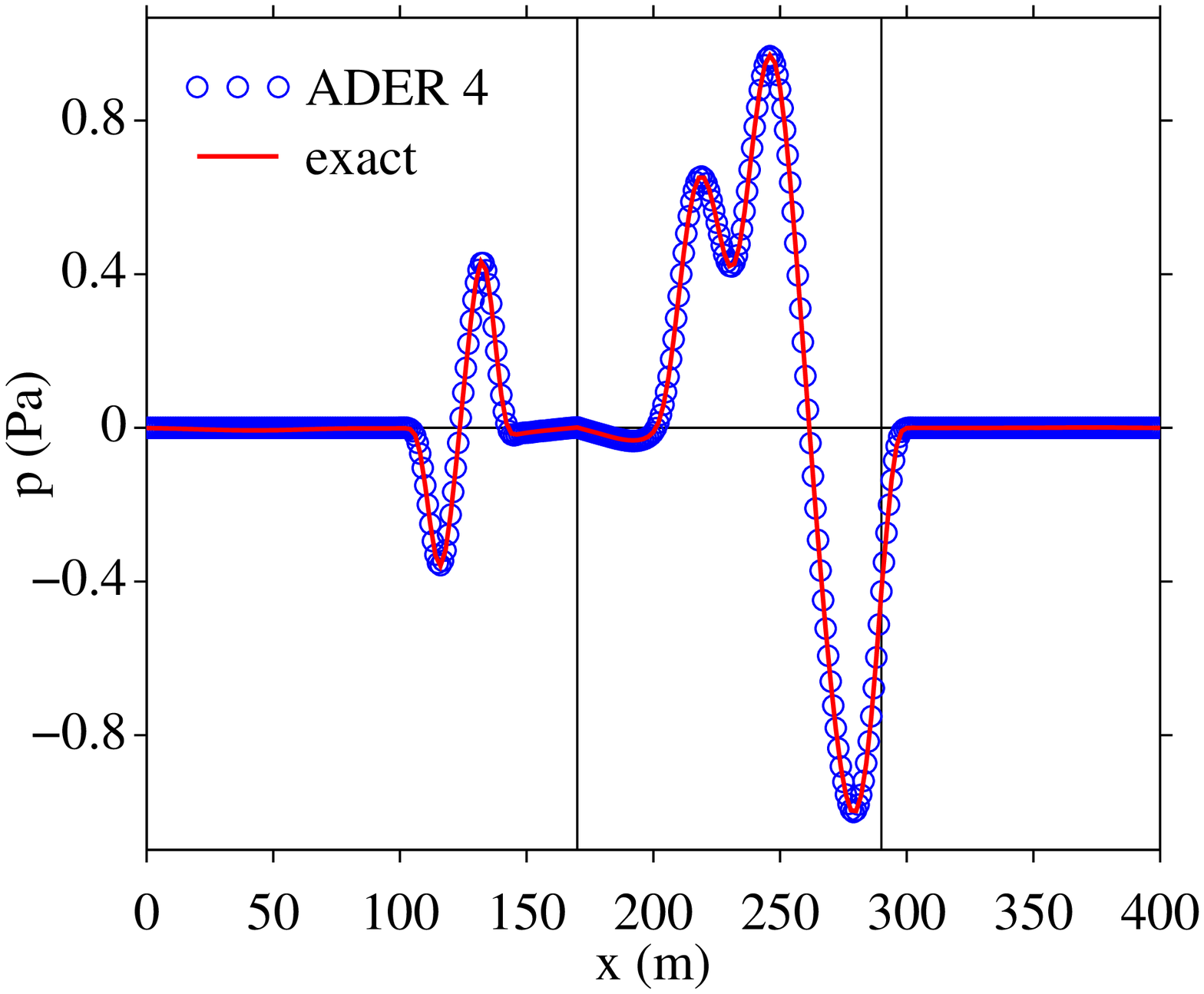}\\
(e) & (f)\\
\includegraphics[scale=0.33]{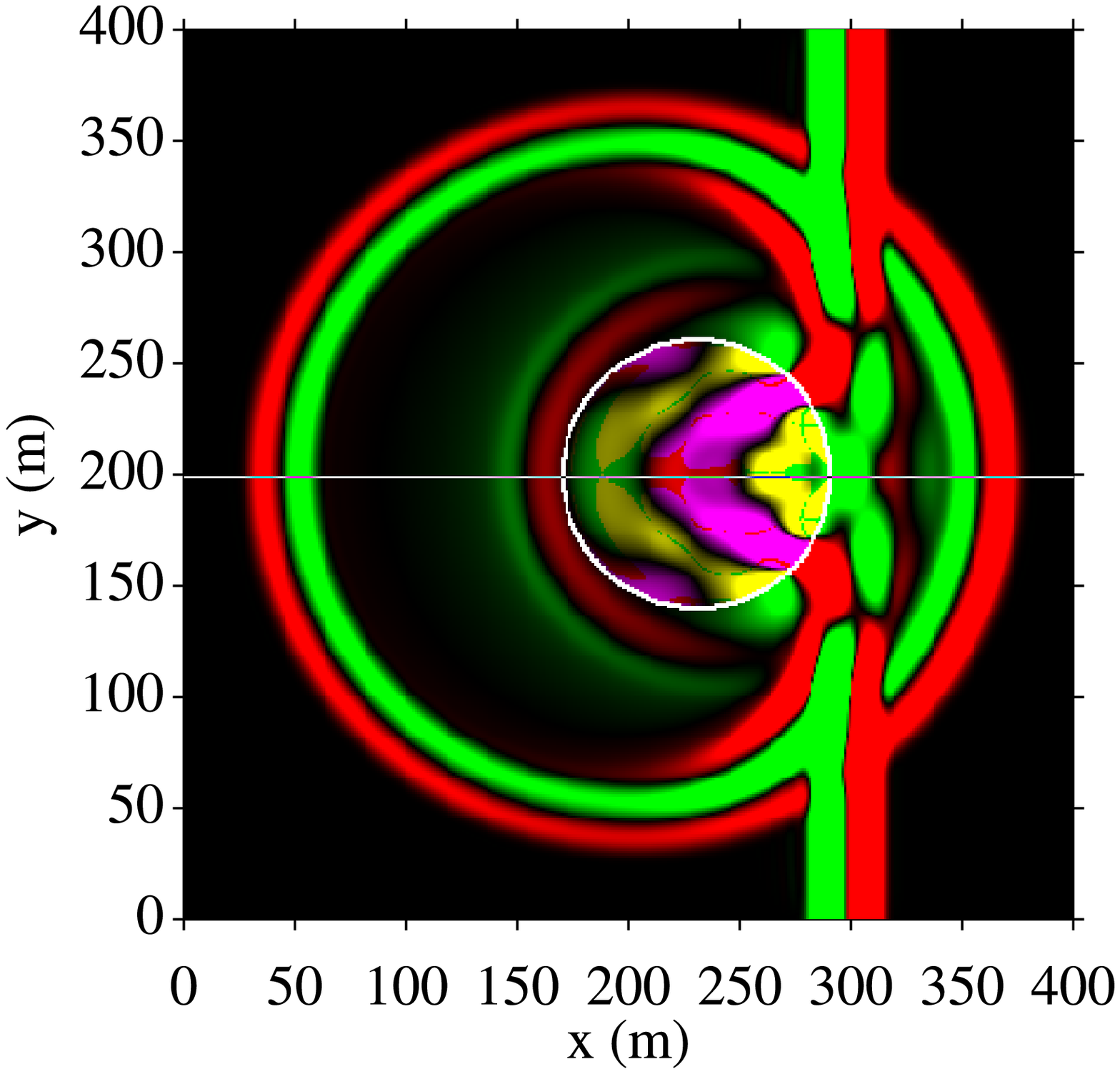}&
\includegraphics[scale=0.33]{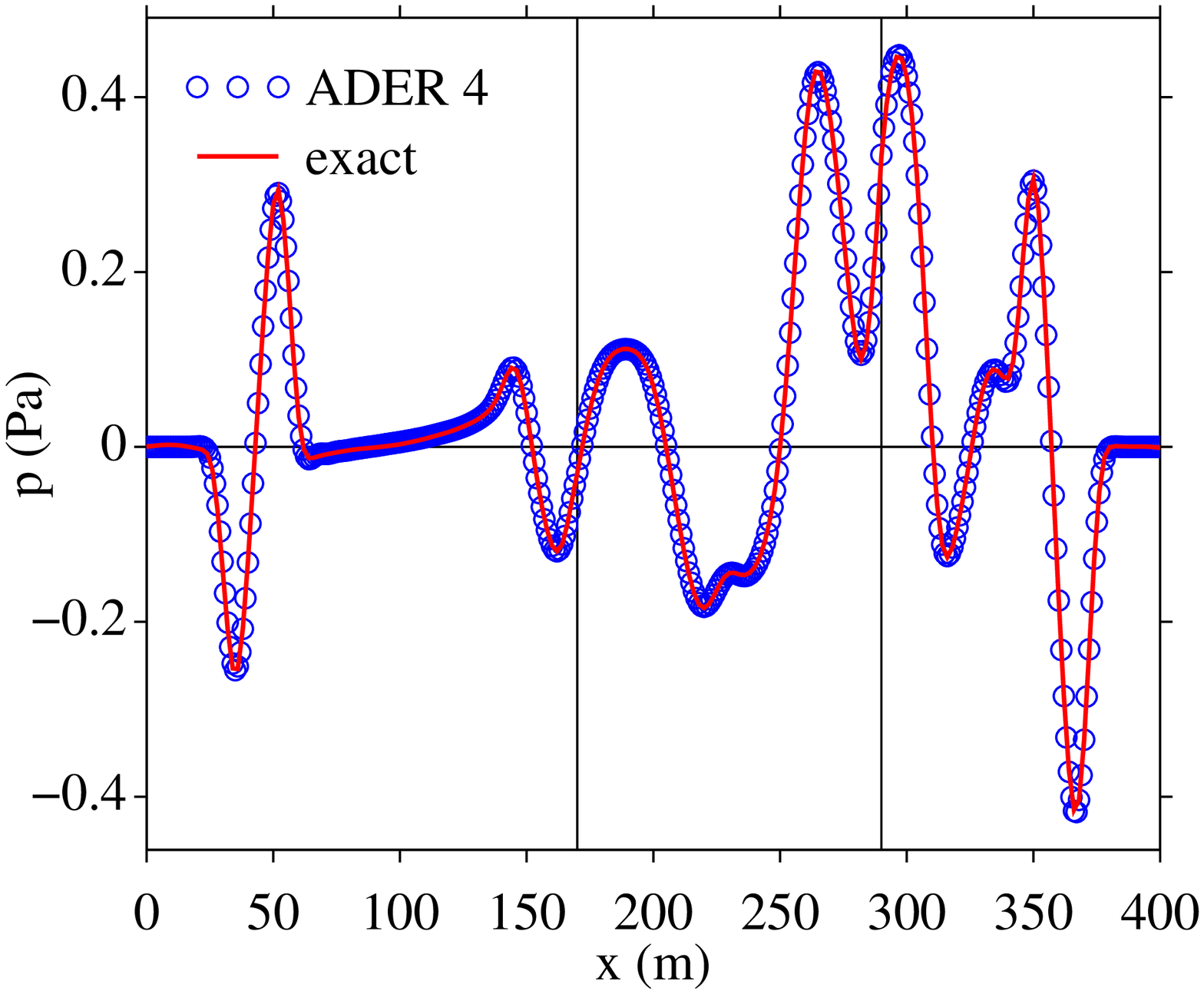}
\end{tabular}
\end{center}
\caption{test 3. Circular interface between a fluid (outside) and a viscoelastic medium (inside). Initial field (a-b), after 220 time steps (c-d) and 440 time steps (e-f).}
\label{FigTest2}
\end{figure}

\begin{figure}[h!]
\begin{center}
\begin{tabular}{cc}
(a) & (b)\\
\includegraphics[scale=0.44]{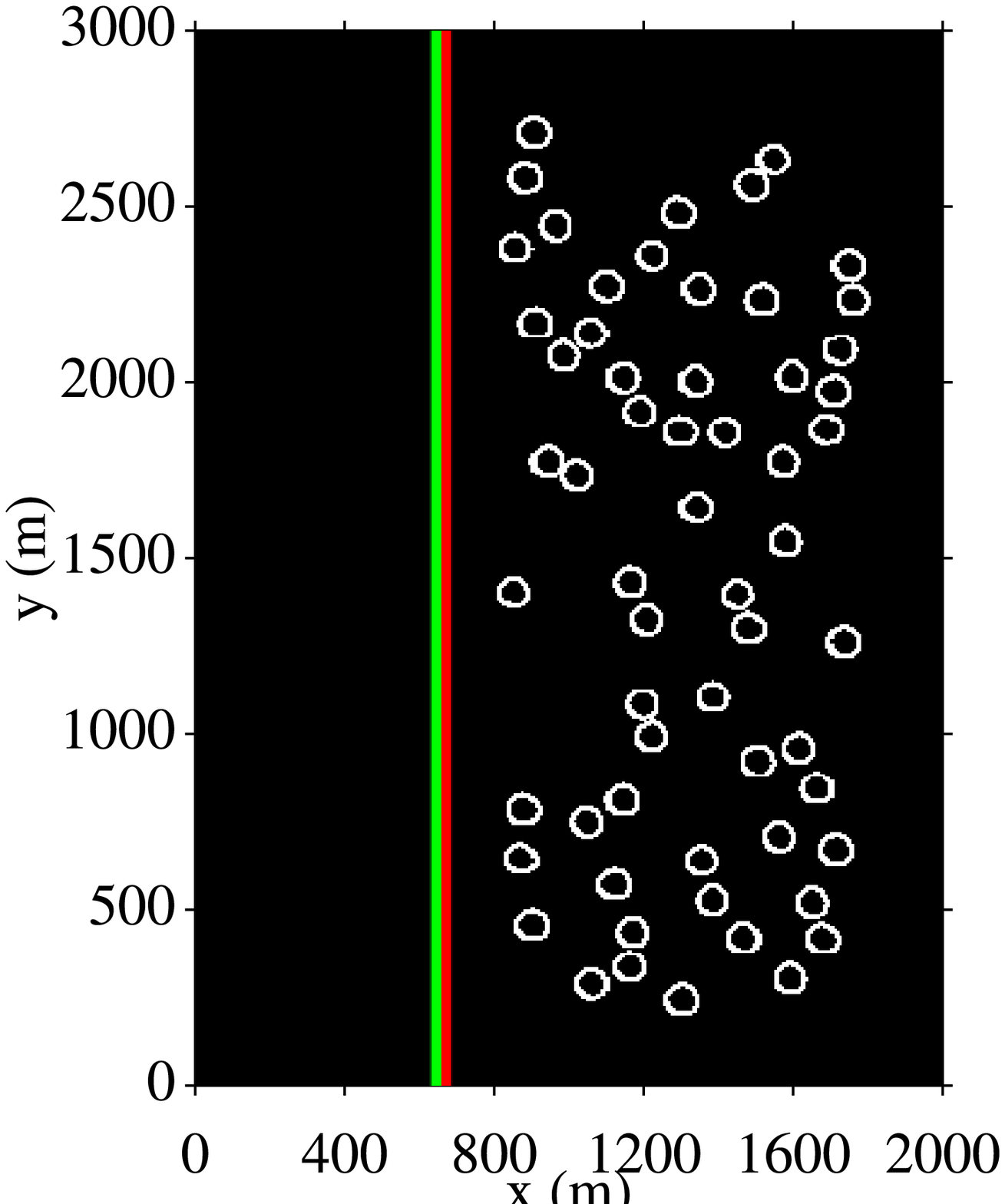}&
\includegraphics[scale=0.44]{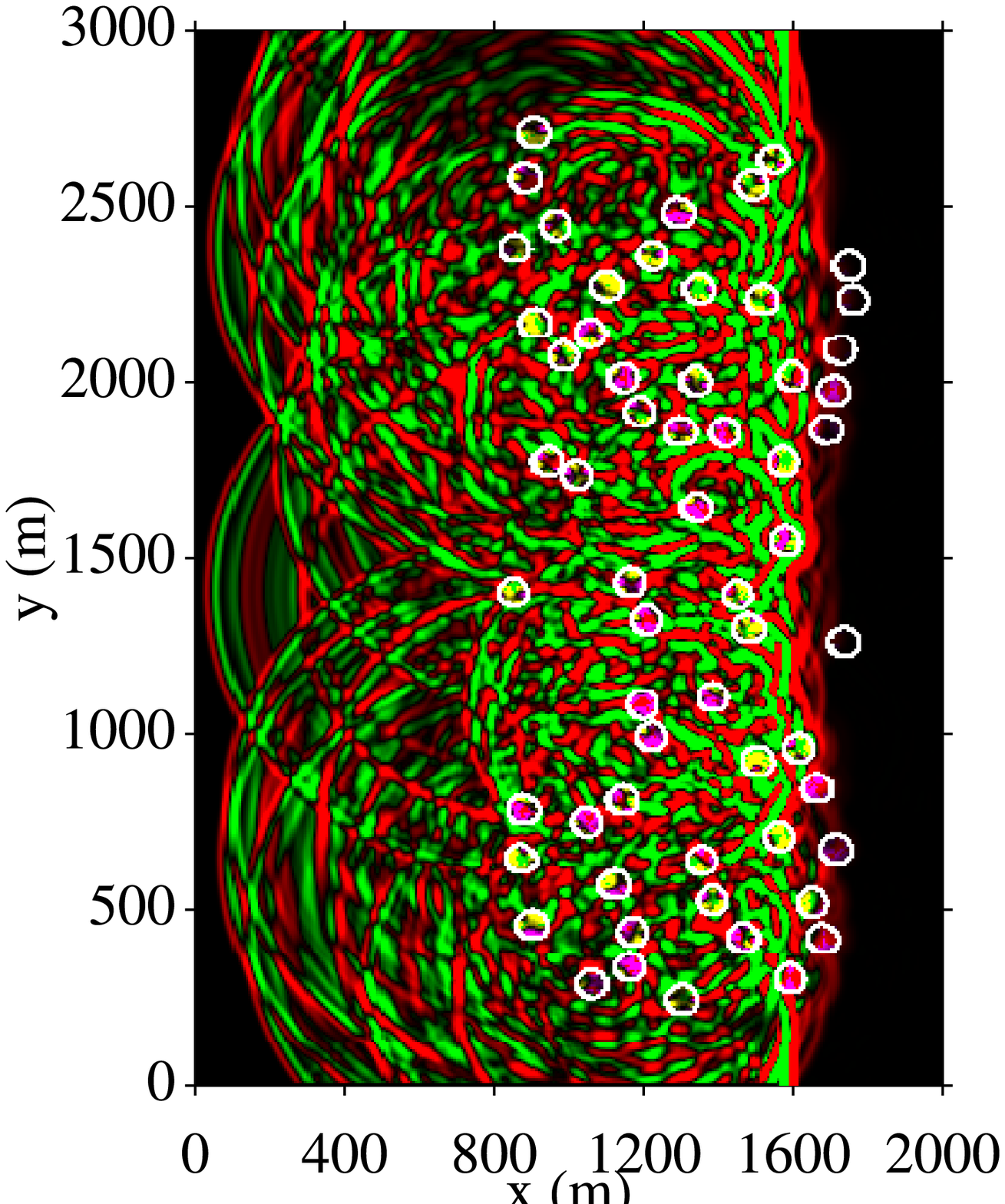}
\end{tabular}
\end{center}
\caption{test 4. Initial (a) and scattered field after 2200 time steps (b).}
\label{FigTest3}
\end{figure}

The immersed interface method depends on the curvature of the interface and its successive derivatives \cite{LOMBARD04,LOMBARD06}. To test the method with a non-null curvature, we now examine a circular interface with a radius of 60 m. The fluid and the viscoelastic medium are outside and inside the circle, respectively. The incident field is a plane acoustic wave with a horizontal wave vector. Figure \ref{FigTest2} shows the field at the initial instant (a-b), after 220 time steps (c-d) and after 440 time steps (e-f). Classical conversions and scattering phenomena are observed. Excellent agreement is observed between the numerical and the exact values. The latter are computed using Fourier techniques and decomposing the plane waves on the basis of Bessel functions.


\subsection{Test 4: plane wave and multiple scattering in random medium}\label{SecNumScatt}

In the previous tests, the validity of the numerical scheme and the immersed interface method was confirmed in the case of academic configurations. We now take a complex medium composed of 60 viscoelastic cylinders randomly embedded in water. The computations are performed on $2000 \times 3000$ grid nodes. Figure \ref{FigTest3} shows the initial field (a) and the scattered fields after 2200 time steps (b), when the incident wave has propagated over a distance corresponding to 12 wavelengths.


\section{Perspectives}\label{SecPerspective}

The propagation of mechanical waves in dissipative solids was addressed numerically in the time domain. To avoid dealing with convolution products, memory variables were introduced. Evolution equations were splitted into two parts: the propagative part was solved numerically using a fourth-order scheme for hyperbolic systems; and the diffusive part was solved exactly. The jump conditions were discretized by means of an immersed interface method, which introduced a subcell resolution on a Cartesian grid. In numerical experiments, focus was put on the fluid / viscoelastic interface, but the algorithms have been implemented and tested in many other cases, such as viscoelastic / vacuum and viscoelastic / viscoelastic interfaces.

The numerical methods presented here make it possible to simulate physically relevant numerical experiments, for instance multiple scattering in random media as performed in section \ref{SecNumScatt}. By applying signal processing tools on the simulated data, it is possible to determine the properties of the {\it effective medium} which is equivalent to the disordered medium investigated \cite{CHEKROUN09}. This numerical approach can be used advantageously instead of the methods usually adopted by physicists so far: real experiments are expensive, and analytical methods can be used only with very small concentrations of scatterers. The latter limitation is particularly penalizing in the case of concrete, where the concentration of aggregates lies typically around 40\%.


\section*{Acknowledgments}

We are grateful to the reviewers for their useful comments and to have suggested some improvements to this manuscript. Many thanks also to Jessica Blanc for her careful reading.


\appendix

\section{Coefficients of splitting}\label{SecCoeffSplitting}

The coefficients $c_m$ and $d_m$ involved in (\ref{AlgoSplitting1}) and (\ref{AlgoSplitting2}) are detailed up to ${\cal N}$. They satisfy 
$$
\sum_{m=1}^{\cal N}c_m=1,\qquad \sum_{m=1}^{\cal N}d_m=1.
$$
For ${\cal N}=1$, one has the usual coefficients
\begin{equation}
\begin{array}{ll}
c_1=1, & d_1=1.
\end{array}
\label{LieTrotter}
\end{equation}
For ${\cal N}=2$, the classical second-order Strang's splitting is recovered \cite{LEVEQUE02}
\begin{equation}
\begin{array}{ll}
c_1=0, & d_1=1/2,\\
[6pt]
c_2=1, & d_2=1/2.
\end{array}
\label{Strang}
\end{equation}
Since $c_1=0$ in (\ref{Strang}), the first spatial integration in (\ref{AlgoSplitting1}) is not performed. For ${\cal N}=3$, the set of coefficients is \cite{FOREST90}
\begin{equation}
\begin{array}{ll}
c_1=7/24,  & d_1=2/3,\\
[6pt]
c_2=3/4,   & d_2=-2/3,\\
[6pt]
c_3=-1/24, & d_3=1.
\end{array}
\label{Forest3}
\end{equation}
Lastly, setting $\chi=(2^{1/3}+2^{-1/3}-1)/6\approx 0.1756$, the coefficients of fourth-order splitting are \cite{FOREST90}
\begin{equation}
\begin{array}{ll}
c_1=\chi+1/2,   & d_1=0,         \\
[6pt]
c_2=-\chi   ,   & d_2=2\,\chi+1, \\
[6pt]
c_3=-\chi   ,   & d_3=-4\,\chi-1,\\
[6pt]
c_4=\chi+1/2 ,  & d_4=2\,\chi+1.
\end{array}
\label{Ruth4}
\end{equation}


\section{Four steps to build ${\bf U}_{I,J}^*$ (\ref{UIJ*})}\label{SecDetailsEsim}

{\bf Step 1: high-order interface conditions}. First, we seek the interface conditions satisfied by the spatial derivatives of the velocity and stress components at $P$. For this purpose, the zero-th order interface conditions (\ref{JC0}) are differentiated in terms of $t$. The time derivatives are replaced by spatial derivatives, using the propagative part (\ref{SplittingLC})-$(a)$. Equations (\ref{JC0}) are also differentiated in terms of $\tau$, using the chain-rule. For instance, the boundary condition ${\bf L}_0^0\,{\bf U}_0^0={\bf 0}$ results in
\begin{equation}
\begin{array}{lll}
\displaystyle
\frac{\textstyle \partial}{\textstyle \partial\,t}\,({\bf L}_0^0\,{\bf U}_0^0)
=
-{\bf L}_0^0\,\overline{\bf A}_0\,\frac{\textstyle \partial}{\textstyle \partial\,x}\,{\bf U}_0^0-{\bf L}_0^0\,\overline{\bf B}_0\,\frac{\textstyle \partial}{\textstyle \partial\,y}\,{\bf U}_0^0={\bf 0},\\
[10pt]
\displaystyle
\frac{\textstyle \partial}{\textstyle \partial\,\tau}\,({\bf L}_0^0\,{\bf U}_0^0)
=
\displaystyle\left(\frac{\textstyle d}{\textstyle d\,\tau}\,{\bf L}_0^0\right)\,{\bf U}_0^0+{\bf L}_0^0\left(x^{'}\frac{\textstyle \partial}{\textstyle \partial\,x}\,{\bf U}_0^0+y^{'}\frac{\textstyle \partial}{\textstyle \partial\,y}\,{\bf U}_0^0\right)={\bf 0}.
\end{array}
\label{JC1}
\end{equation}
From (\ref{JC1}), a matrix ${\bf L}_0^1$ is built such that ${\bf L}_0^1\,{\bf U}_0^1={\bf 0}$. This matrix depends on $\tau$ and on the physical parameters on $\Omega_0$. Applying a similar procedure to the three equations in (\ref{JC0}) gives a set of first-order interface conditions. By iterating this process $k$ times, we obtain the $k$-th order interface conditions
\begin{equation}
{\bf C}_1^k\,{\bf U}_1^k={\bf C}_0^k\,{\bf U}_0^k,\qquad
{\bf L}_\ell^k\,{\bf U}_\ell^k={\bf 0},\qquad \ell=0,1.
\label{JCk}
\end{equation}
When $k\geq 2$, building the matrices ${\bf C}_\ell^k$ and ${\bf L}_\ell^k$ is a tedious task, which can be greatly simplified by using computer algebra tools. Note lastly that ${\bf C}_\ell^k$ and ${\bf L}_\ell^k$ involve the spatial derivatives $\frac{d^\alpha\,x}{d\,\tau^\alpha}$ and $\frac{d^\alpha\,y}{d\,\tau^\alpha}$ ($\alpha=1,...,\,k+1$), which provides insights about the local geometry of $\Gamma$ at $P$. \\

{\bf Step 2: high-order Beltrami equations}. In (\ref{SplittingLC})-$(a)$, the viscoelastic medium behaves like an elastic medium with Lam\'e coefficients $\lambda=\pi_u-2\,\mu_u$ and $\mu=\mu_u$, where compatibility conditions are satisfied between some spatial derivatives of the strain components \cite{LOVE44}. When expressed in terms of ${\bf \sigma}$, these conditions lead to the Beltrami equation
\begin{equation}
\frac{\textstyle \partial^2}{\textstyle \partial\,x\,\partial\,y}\,\sigma_{12}= \alpha_2\,\frac{\textstyle \partial^2}{\textstyle \partial\,x^2}\,\sigma_{11}+\alpha_1\,\frac{\textstyle \partial^2}{\textstyle \partial\,x^2}\,\sigma_{22}+\alpha_1\,\frac{\textstyle \partial^2}{\textstyle \partial\,y^2}\,\sigma_{11}+\alpha_2\,\frac{\textstyle \partial^2}{\textstyle \partial\,y^2}\,\sigma_{22},   
\label{Compatibilite}
\end{equation}
where
\begin{equation}
\begin{array}{l}
\displaystyle
\alpha_1=\frac{\textstyle \pi_u}{\textstyle 4\,(\pi_u-\mu_u)}=\frac{\textstyle c_{p_\infty}^2}{\textstyle 4\,\left(c_{p_\infty}^2-c_{s_\infty}^2\right)},\\
[14pt]
\displaystyle
\alpha_2=-\frac{\textstyle \pi_u-2\,\mu_u}{\textstyle 4\,(\pi_u-\mu_u)}=\frac{\textstyle 2\,c_{s_\infty}^2-c_{p_\infty}^2}{\textstyle 4\,\left(c_{p_\infty}^2-c_{s_\infty}^2\right)}.
\end{array}
\label{Alp1Alp2}
\end{equation}

The equation (\ref{Compatibilite}) is satisfied anywhere in $\Omega_0$. Under suitable smoothness requirements, it can be differentiated as many times as necessary, with respect to $x$ and $y$. Since the equations thus obtained are also valid along $\Gamma$, they can be used to obtain a minimum number of independent components ${\bf V}_\ell^k$ 
\begin{equation}
{\bf U}_\ell^k={\bf G}_\ell^k\,{\bf V}_\ell^k,\qquad \ell=0,\,1.
\label{MatG}
\end{equation}
The algorithm for building the matrices ${\bf G}_\ell^k$ presented in \cite{LOMBARD06} can be easily adapted to (\ref{Compatibilite})-(\ref{Alp1Alp2}). If $\Omega_1$ is not a viscoelastic medium, then (\ref{MatG}) is still valid if appropriate Beltrami-like equations are used: see \cite{LOMBARD04} for the fluid-elastic case.\\

{\bf Step 3: high-order boundary values}. The high-order boundary conditions in (\ref{JCk}) and the high-order Beltrami equations (\ref{MatG}) give the underdetermined linear systems
\begin{equation}
{\bf L}_\ell^k\,{\bf G}_\ell^k\,{\bf V}_\ell^k={\bf 0},\quad \ell=0,\,1.
\end{equation}
We obtain
\begin{equation}
{\bf V}_\ell^k={\bf K}_\ell^k\,{\bf W}_\ell^k,\quad \ell=0,\,1,
\label{VKW}
\end{equation}
where ${\bf K}_\ell^k$ are the matrices built from the kernel of ${\bf L}_\ell^k\,{\bf G}_\ell^k$. The solution ${\bf W}_\ell^k$ is the minimum set of independent components of the trace of ${\bf U}$ and its spatial derivatives up to the $k$-th order, on the side $\Omega_\ell$. Injecting (\ref{VKW}) into the high-order jump conditions (\ref{JCk}) gives
\begin{equation}
{\bf S}_1^k\,{\bf W}_1^k={\bf S}_0^k\,{\bf W}_0^k ,
\label{CGK}
\end{equation}
where ${\bf S}_\ell^k={\bf C}_\ell^k\,{\bf G}_\ell^k\,{\bf K}_\ell^k$. The underdetermined system (\ref{CGK}) is solved
\begin{equation}
{\bf W}_1^k=\left(\left({\bf S}_1^k\right)^{-1}\,{\bf S}_0^k\,|\,{\bf R}_{{\bf S}_1^k}\right)
\left(
\begin{array}{c}
\displaystyle
{\bf W}_0^k\\
[8pt]
\displaystyle
{\bf \Lambda}^k
\end{array}
\right),
\label{SVD}
\end{equation}
where $({\bf S}_1^k)^{-1}$ is the least-squares pseudo-inverse of ${\bf S}_1^k$, ${\bf R}_{{\bf S}_1^k}$ is the matrix containing the kernel of ${\bf S}_1^k$, and ${\bf \Lambda}^k$ is a set of Lagrange multipliers. To build $({\bf S}_1^k)^{-1}$ and ${\bf R}_{{\bf S}_1^k}$, a singular value decomposition of ${\bf S}_1^k$ is performed.\\

{\bf Step 4: construction of modified values}. We assume that ${\bf U}_{i,j}^{(2m-2)}$ values are known in the splitting algorithm (\ref{AlgoSplitting1}), and hence the restrictions $\overline{{\bf U}}_{i,j}^{(2m-2)}$ are also known. Our goal is to determine a modified solution ${\bf U}_{I,J}^*$ at this time step, to be injected in the discrete operator ${\bf H}_a$. For this purpose, we introduce some notations. Let $P$ be the orthogonal projection of $(x_I,\,y_J)$ on $\Gamma$ (figure \ref{Patate}). The coefficients of 2-D Taylor expansions around $P$ are put in the matrix ${\bf \Pi}_{i,j}^k$: 
\begin{equation}
{\bf \Pi}_{i,j}^k=\left({\bf I},...,
\frac{\textstyle 1}{\textstyle \beta \,!\,(\alpha-\beta)\,!}\,(x_i-x_P)^{\alpha-\beta}(y_j-y_P)^\beta\,{\bf I},...,
\frac{\textstyle (y_j-y_P)^k}{\textstyle k\,!}\,{\bf I}
\right),
\label{Taylor}
\end{equation}
where $\alpha=0,...,\,k$ and $\beta=0,...,\,\alpha$; ${\bf I}$ is the $3\times3$ or $5\times5$ identity matrix, depending on whether $(x_i,\,y_j)$ belongs to a fluid or a viscoelastic medium. By definition, the modified value at $(x_I,\,y_J)$ is
\begin{equation}
{\bf U}_{I,J}^*={\bf \Pi}_{I,J}^k\,{\bf U}_0^k.
\label{SolModa}
\end{equation}
The trace ${\bf U}_0^k$ in (\ref{SolModa}) still remains to be determined in terms of the interface conditions and the numerical values ${\bf U}_{i,j}^{(2m-2)}$ at surrounding nodes. 

Consider the disc ${\cal D}$ is centered at $P$ with radius $q$ (figure \ref{Patate}). At the grid nodes of ${\cal D}\cap\Omega_0$, $k$-th order Taylor expansion of the solution at $P$, and the conditions (\ref{MatG}) and (\ref{VKW}), give
\begin{equation}
\begin{array}{lll}
\overline{\bf U}_{i,j}^{(2m-2)} &=& {\bf \Pi}_{i,j}^k\,{\bf U}_0^k,\\
&&\\
&=& {\bf \Pi}_{i,j}^k\,{\bf G}_0^k\,{\bf K}_0^k\,{\bf W}_0^k,\\
[10pt]
&=& {\bf \Pi}_{i,j}^k\,{\bf G}_0^k\,{\bf K}_0^k\,\left({\bf 1}\,|\,{\bf 0}\right)
\left(
\begin{array}{c}
{\bf W}_0^k\\
\\
{\bf \Lambda}^k
\end{array}
\right)
.
\end{array}
\label{Taylor1}
\end{equation}
At the grid nodes of ${\cal D}\cap\Omega_1$, $k$-th order Taylor expansion of the solution at $P$, and the interface conditions (\ref{MatG}), (\ref{VKW}) and (\ref{SVD}), give 
\begin{equation}
\begin{array}{lll}
\overline{\bf U}_{i,j}^{(2m-2)} &=& {\bf \Pi}_{i,j}^k\,{\bf U}_1^k,\\
&&\\
&=& {\bf \Pi}_{i,j}^k\,{\bf G}_1^k\,{\bf K}_1^k\,{\bf W}_1^k,\\
[10pt]
&=& {\bf \Pi}_{i,j}^k\,{\bf G}_1^k\,{\bf K}_1^k\,\left(\left({\bf S}_1^k\right)^{-1}\,|\,{\bf R}_{{\bf S}_1^k}\right)
\left(
\begin{array}{c}
{\bf W}_0^k\\
\\
{\bf \Lambda}^k
\end{array}
\right)
.
\end{array}
\label{Taylor2}
\end{equation}
The equations (\ref{Taylor1}) and (\ref{Taylor2}) are written using an adequate matrix ${\bf M}$
\begin{equation}
\left(
\overline{\bf U}^{(2m-2)}
\right)_{\mathcal D}
={\bf M}
\left(
\begin{array}{c}
{\bf W}_0^k\\
\\
{\bf \Lambda}^k
\end{array}
\right).
\label{Taylor3}
\end{equation}
The radius $q$ is chosen so that (\ref{Taylor3}) is overdetermined. 
The least-squares inverse of the matrix ${\bf M}$ is denoted by ${\bf M}^{-1}$. Since the Lagrange multipliers ${\bf \Lambda}^k$ are not involved in (\ref{SolModa}), ${\bf M}^{-1}$ is restricted to $\overline{\bf M}^{-1}$, so that
\begin{equation}
{\bf W}_0^k=\overline{\bf M}^{-1}
\,\left(
\overline{\bf U}^{(2m-2)}
\right)_{\mathcal D}.
\label{Taylor4}
\end{equation}
The modified value follows from (\ref{MatG}), (\ref{VKW}), (\ref{SolModa}) and (\ref{Taylor4}):
\begin{equation}
\begin{array}{lll}
\displaystyle
{\bf U}_{I,J}^* &=& \displaystyle {\bf \Pi}_{I,J}^k\,{\bf G}_0^k\,{\bf K}_0^k\,\overline{\bf M}^{-1}\,\left(\overline{\bf U}^{(2m-2)}\right)_{\mathcal D},\\
[12pt]
&=& \displaystyle {\cal M}\,\left(\overline{\bf U}^{(2m-2)}\right)_{\mathcal D}.
\end{array}
\label{Esim}
\end{equation}


\end{document}